\documentclass[aps,prb,twocolumn,amsmath,superscriptaddress,amssymb,floatfix]{revtex4}

\usepackage{epsf}
\usepackage{natbib}
\setcitestyle{square,numbers}

\usepackage{tabularx} 
\usepackage{graphicx} 
\usepackage{epstopdf}
\usepackage{epsf}
\usepackage{dcolumn}
\usepackage{bm}
\usepackage{color,xcolor}
\usepackage[colorlinks=true,citecolor=blue, linkcolor=red]{hyperref}

\usepackage{multirow}
\usepackage[normalem]{ulem}
\usepackage{color,soul}
\setul{0.5ex}{0.5ex}
\setulcolor{red}

\begin{document}

\title{Orbital-Induced Peierls Transitions: How Orbitals Orchestrate Lattice Instability}
  
\author{Takashi Mizokawa}
\thanks{Corresponding author. Email: mizokawa@waseda.jp}
\affiliation{Department of Applied Physics, Waseda University, Shinjuku, Tokyo 169-8555, Japan}

\author{Sergey V. Streltsov}
\thanks{Corresponding author. Email: streltsov.s@gmail.com}
\affiliation{Institute of Metal Physics, Ural Branch of the Russian Academy of Sciences, Ekaterinburg 620137, Russia}

\date{\today}

\begin{abstract}
The Peierls transition is typically regarded as a phenomenon inherent to one-dimensional (1D) materials. However, orbital degrees of freedom can induce this instability even in higher dimensions. Two mechanisms are primarily responsible. First, the anisotropic shape of $p$ and $d$ orbitals can lead to effective ``1D-zation'' of the electronic spectrum. Second, orbital degrees of freedom can lift band degeneracy by shifting bands relative to each other via the local or band Jahn–Teller effect, thereby affecting the nesting of the Fermi surface. The orbital-induced Peierls effect is most commonly observed when ligand octahedra surrounding transition metals share edges, and less frequently in face-sharing geometries. In this review, we discuss the underlying physical mechanisms, the materials in which this phenomenon occurs, the characteristics of the high-temperature undistorted phase, and the role of local effects such as the formation of molecular orbitals.

\end{abstract}

\maketitle

\hfill \break {\bf \noindent INTRODUCTION\\} 
The Peierls transition is undoubtedly one of the most renowned phenomena in condensed matter physics. It was first predicted theoretically~\cite{footnote1} and subsequently confirmed experimentally~\cite{comes1973,monceau1976,dumas1983}.

\begin{figure}[t!]
\centering
\includegraphics[width=0.7\columnwidth]{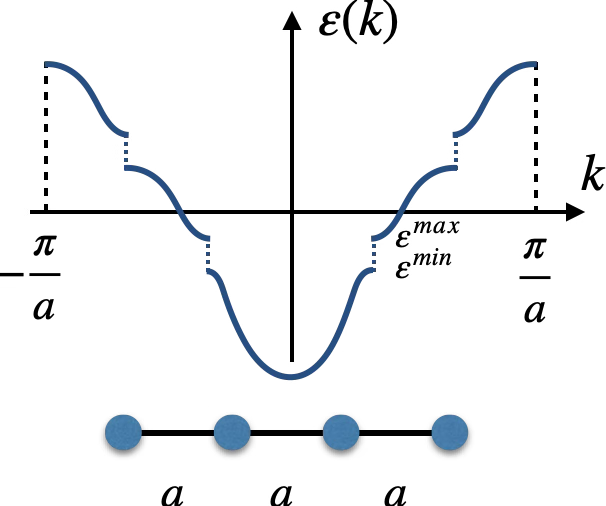}
\caption{
Peierls was the first to show that the electron dispersion of a linear chain of atoms spaced at distance $a$ is modified by lattice distortions. The vertical breaks in the sketch are proportional to corresponding displacements. If the electron filling is such that the Fermi energy lies exactly within the gap, the distortion lowers the electronic energy. Otherwise, the distortion is ineffective: the average of the maximal $\varepsilon^{max}$ and minimal $\varepsilon^{min}$ energies remains equal to the band energy of the undistorted chain, while the elastic energy cost is always positive. Hence, no net gain is achieved.
}
\label{Fig:1D-gaps}
\end{figure}

\begin{figure}[t!]
\centering
\includegraphics[width=1\columnwidth]{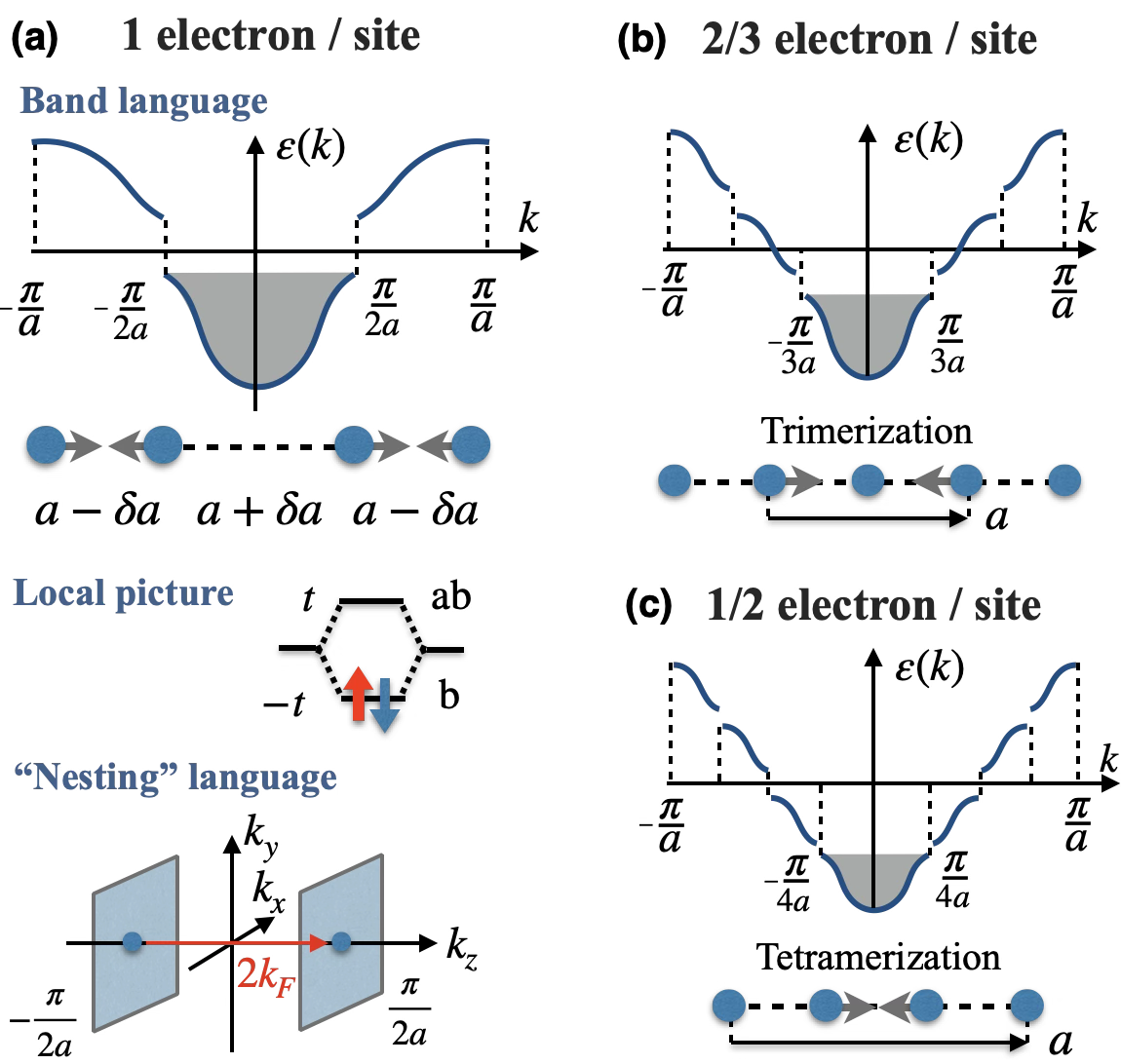}
\caption{Illustration of the Peierls transition for different numbers of electrons (a-c). For the case of a single electron, different interpretations of the transition are presented: the instability leading to gap opening in the electronic spectrum is illustrated in the top panel of (a). The formation of bonding and antibonding states assuming orthogonal wavefunction due to dimerization and corresponding energy gain is shown in the middle panel of (a). The bottom panel of (a) illustrates nesting of the Fermi surface, which leads to divergence of the generalized susceptibility in Lindhard theory or in the random phase approximation (RPA), see Eq.~\eqref{chi0}.}
\label{Fig:1D-diff-fillings}
\end{figure}

Peierls considered a linear chain with distance $a$ between atoms and a one-dimensional (1D) electron gas, and showed that it is always energetically favorable to break the uniform spacing. The resulting lattice distortion leads to the appearance of gaps at wavevectors $k_n = \pm n\pi/ma$, where $n$ and $m$ are  integers and $n=1,2,..., m-1$~\cite{Peierls1996}, see Fig.~\ref{Fig:1D-gaps}. Significant energy is gained if the band is exactly filled up to one of these gaps  at $k_n$. For example, in a chain of $s$ ions at half-filling (one electron per site), such a gap appears precisely at $k_F = \pm \pi/2a$, as shown in the top panel of Fig.~\ref{Fig:1D-diff-fillings}(a). Consequently, the chain becomes unstable toward dimerization, which is responsible for the formation of these gaps in the $\varepsilon(k)$ dispersion. Indeed, one can show that for small distortions $\tilde u$, the gain in electronic energy $\sim \tilde u^2 \ln \tilde u^2$ always dominates over the loss in elastic energy $\sim B \tilde u^2/2$ due to additional lattice distortions~\cite{Khomskii-book1}.

In chemical terms, this is essentially the formation of hydrogen molecules, as depicted in the middle part of Fig.~\ref{Fig:1D-diff-fillings}(a). At half-filling, only bonding levels are occupied, resulting in an energy gain by the formation of molecular orbitals, with $E_{b} = -2t$ per dimer ($t$ is the hopping amplitude between these two sites). In contrast, for two electrons per site, no such gain occurs, since both bonding and antibonding levels are occupied. In that case, the system experiences an energy loss due to elastic deformation, which counteracts dimerization.

The Peierls transition affects the distribution of electrons and is therefore closely connected with the formation of charge density waves (CDWs)~\cite{Pouget2016,Khomskii-book1}. In the simplest case of half-filling, small on-site electron correlations (Hubbard $U$), and a chain geometry, the electronic density is concentrated within the dimer; i.e., a bond-centered CDW is formed in this situation. 

Microscopically, one of the driving forces of a CDW or Peierls transition is the momentum-dependent electron-phonon coupling. In this picture, phonons modulate the hopping integrals $t$, and this turns out to be extremely efficient when, e.g., molecular bonding (in clusters) occurs. Interestingly, one can exclude phonons from the treatment and then end up with an effective electron-electron attraction, which results in CDW formation or the Peierls transition~\cite{Khomskii-book1}.

While the Peierls transition is often referred to as dimerization, Peierls himself clearly pointed out that the instability can lead to very different types of lattice distortions depending on the electron count~\cite{Peierls1996}. For example, it can result in tetramerization at 1/4-filling, (with gap appearing at $\pm \pi/4a$) or trimerization at 1/3 band filling (with gap at $\pm \pi/3a$), as illustrated in Fig.~\ref{Fig:1D-diff-fillings}(b,c). On the other hand, increasing the dimensionality of the system - by going to two-dimensional (2D) or three-dimensional (3D) structures - inevitably introduces band dispersion in the additional dimensions. This tends to close the aforementioned gaps and thereby minimize or even completely eliminate the energy gain associated with the lattice period modulation that is so effective in 1D.

In more modern terms, the Peierls transition is associated with nesting of the Fermi surface. For a chain aligned along the $z$-direction, the Fermi surface consists of two planes perpendicular to $k_z$, as shown in the lowest part of Fig.~\ref{Fig:1D-diff-fillings}(a). They can be matched by shifting one of them by a vector $\vec q = 2\vec k_F$. The generalized susceptibility of the non-interacting electron system in the static limit is given by the Lindhard formula:
\begin{eqnarray}
\label{chi0}
\chi_0(\vec q) =
 \sum_{\vec  k} \sum_{m m'} \frac{n_m(\vec k + \vec q) - n_{m'}(\vec k)} {\varepsilon_{m} ({\vec  k+ \vec q}) -   \varepsilon_{m'}({\vec k}) - i \theta},
\end{eqnarray}
where $m$ and $m'$ are band indexes $n(\vec k)$ is the Fermi–Dirac distribution function and $\theta$ is a positive infinitesimal that ensures the correct causal behavior, see e.g. \cite{Jerome1982,fazekas}. 

Shifting one of the planes that form the Fermi surface by a wavevector $\vec q$ leads to a divergence in the generalized susceptibility — a hallmark of instability (Fermi surface nesting~\cite{fazekas}, see also Fig.~\ref{Fig:1D-diff-fillings}(a) bottom panel). 
When going from one to two or three dimensions, the Fermi surface often deforms due to hybridization (hopping) along the extra directions and may even lose its nesting condition.

The key ingredient that helps to extend this essentially 1D physics to higher dimensionality is the orbital degrees of freedom. Both $p$ and $d$ orbitals have a very anisotropic spatial distribution, which facilitates electron hopping in one direction but nearly completely suppresses it in others. 

{\it Thus, the first effect related to orbital degrees of freedom is that their anisotropic shape can result in ``1D-zation'' of the electronic band structure~\cite{Streltsov2017}, and this promotes the Peierls transition.}

\begin{figure}[t!]
\centering
\includegraphics[width=1\columnwidth]{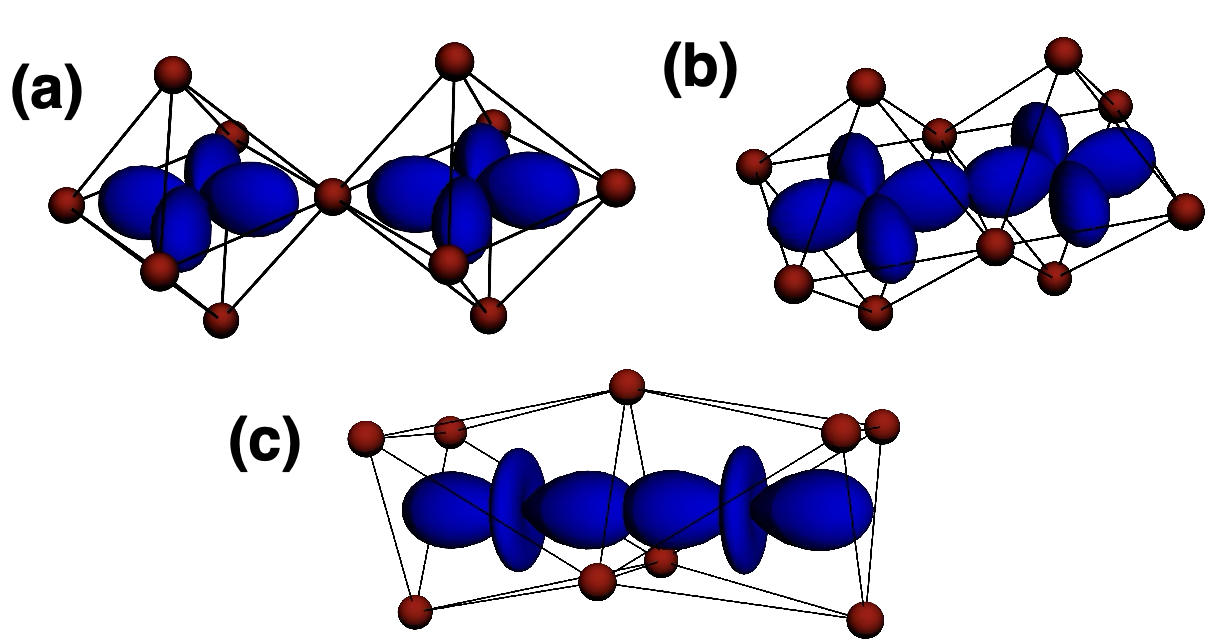}
\caption{Different type of sharing in octahedral geometry: (a) common corner with the $x^2-y^2$ orbital, (b) common edge with the $xy$ orbital, and (c) common face with the $a_{1g}$ orbital.}
\label{Fig:sharing}
\end{figure}

In the present review, we restrict ourselves to considering only $d$ orbitals. In octahedral geometry, the $e_g$ orbitals are directed toward the ligands, while the $t_{2g}$ point away from them. In a situation, when neighboring $MX_6$ ($M$ = transition metal, $X$=ligands)  octahedra share common corners, the strongest is the $e_g$-$e_g$ hopping between two metals via ligand $p$ orbital directed along this bond, see Fig.~\ref{Fig:sharing}(a). In contrast, when neighboring octahedra share their edges or faces, direct hopping between $t_{2g}$ orbitals becomes extremely efficient, as shown in Fig.~\ref{Fig:sharing}(b) and (c). We will see in what follows that the orbital-induced Peierls transition occurs nearly exclusively in these two geometries and is facilitated by direct $t_{2g}$-$t_{2g}$ hopping (although CDW instabilities are not limited by these cases and can occur in any geometry). The physical origin of this feature is that a ligand lies between two transition metals in the common-corner geometry, preventing them from being brought close enough to achieve maximal energy gain through the formation of molecular orbitals between $d$ orbitals.

{\it The second ingredient that orbitals bring to Peierls physics is the (band) Jahn–Teller effect.}
\begin{figure}[b!]
 \centering
\includegraphics[width=0.7\textwidth]{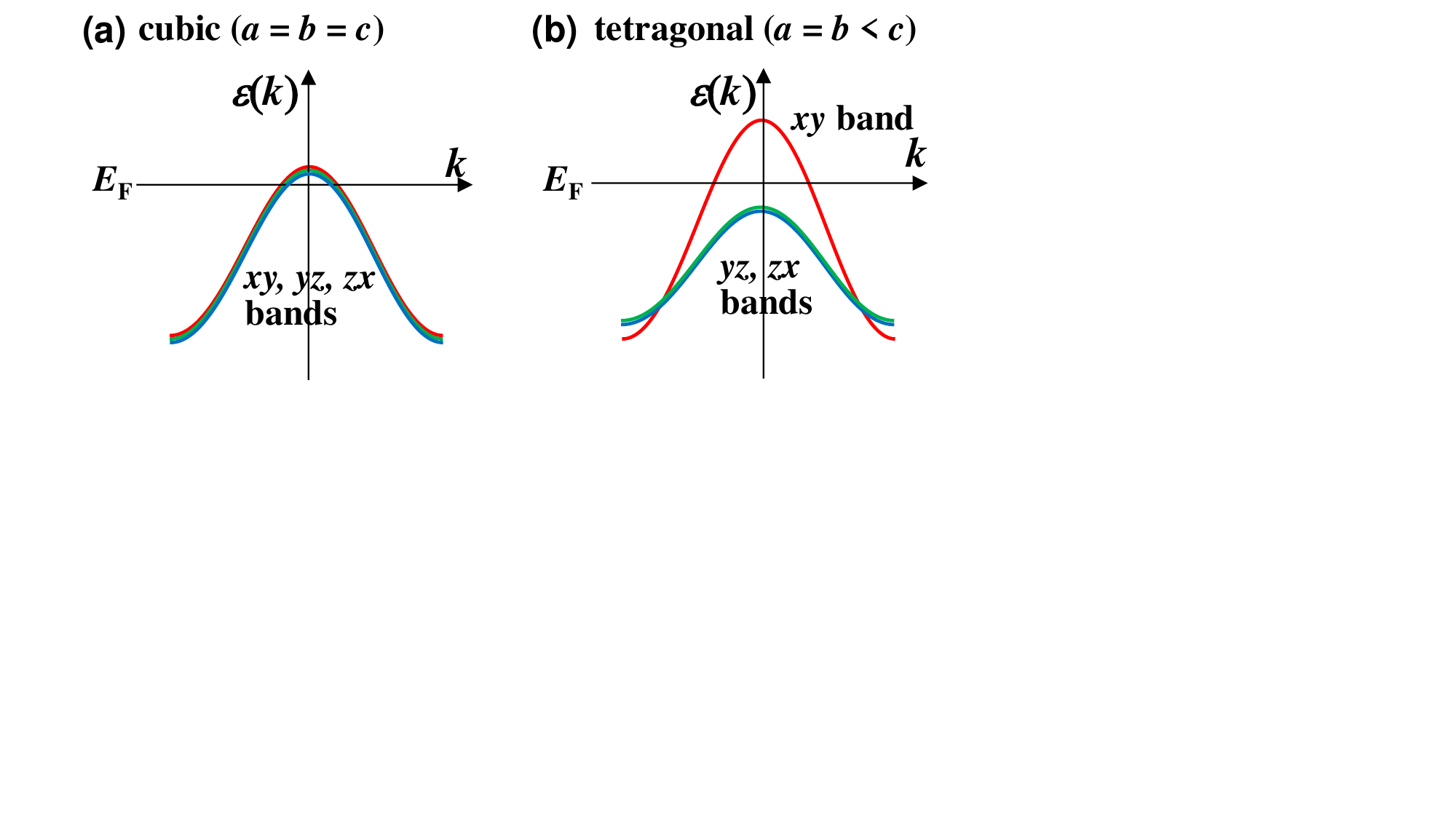}
\vspace{-4cm}
\caption{Schematic drawings for $t_{2g}$ band dispersions of $AB_2X_4$ spinels (transition metals at B sites) in the cubic ($a=b=c$) situation (a) and the elongated tetragonal ($a=b<c$) situation (b). 
Setting the $x$-, $y$-, and $z$-axes along the $a$, $b$, and $c$ directions of the cubic/tetragonal crystal, the $xy$, $yz$, and $zx$ bands run along the [1,$\pm$1,0], [0,1,$\pm$1], and [$\pm$1,0,1] directions, respectively. When the Fermi level is close to the upper bound of the $t_{2g}$ bands, holes can be accommodated only in the $xy$ band due to the band Jahn-Teller distortion with $a=b<c$ along the $c$- or $z$-axis.} 
\label{bJT}
\end{figure}

In the conventional (local) Jahn–Teller effect in materials with localized electrons, vibronic coupling typically results in additional distortions of the ligand polyhedra and a corresponding splitting of energy levels~\cite{Khomskii-book1}. It results in the localization of an electron in a particular orbital and often leads to corresponding orbital ordering and cooperative distortions \cite{KugelKhomskii1982}.

In metals the band Jahn–Teller effect is related not only to the lifting of band degeneracy, but also to changes in the widths of the different bands. As will be illustrated with many examples below, this helps to improve the nesting condition. For instance, in normal $AB_2$X$_4$ spinels, transition metals occupying $B$ sites reside in octahedra that share their edges, and the overall crystal structure is rather 3D and uniform. This implies the degeneracy of the $t_{2g}$ bands, see Fig.~\ref{bJT}(a). However, in CuIr$_2$S$_4$ and LiRh$_2$O$_4$, one of the $t_{2g}$ bands ($xy$ band) becomes wider with the tetragonal elongation along the $c$-axis and the concomitant compression in the $ab$-plane. As a result, the Fermi surfaces are created solely by the $xy$ band rather than the three, see Fig.~\ref{bJT}(b), and the degree of the Fermi surface nesting can be enhanced under the band Jahn-Teller effect. In fact, the band Jahn–Teller effect and the Peierls instability are often intrinsically synergetic: both tend to lower the energy through lattice distortions, and in many systems, tetragonal and Peierls-type distortions set in concurrently at the same transition temperature. Consequently, in real materials, it becomes challenging to unambiguously separate the energy gains arising from the band Jahn–Teller effect from those due to the Peierls mechanism, a point to which we return in the discussion below.

The idea of orbital-induced Peierls mechanism has been introduced in \cite{Khomskii2005a} motivated by this peculiar phase transitions in spinels and then applied to many other classes of materials. In Table~\ref{Tab:Materials} we summarize materials, which low-temperature structures can be explained by the orbital-induced Peierls transition.

\begin{figure}[t!]
 \centering
\includegraphics[width=0.5\textwidth]{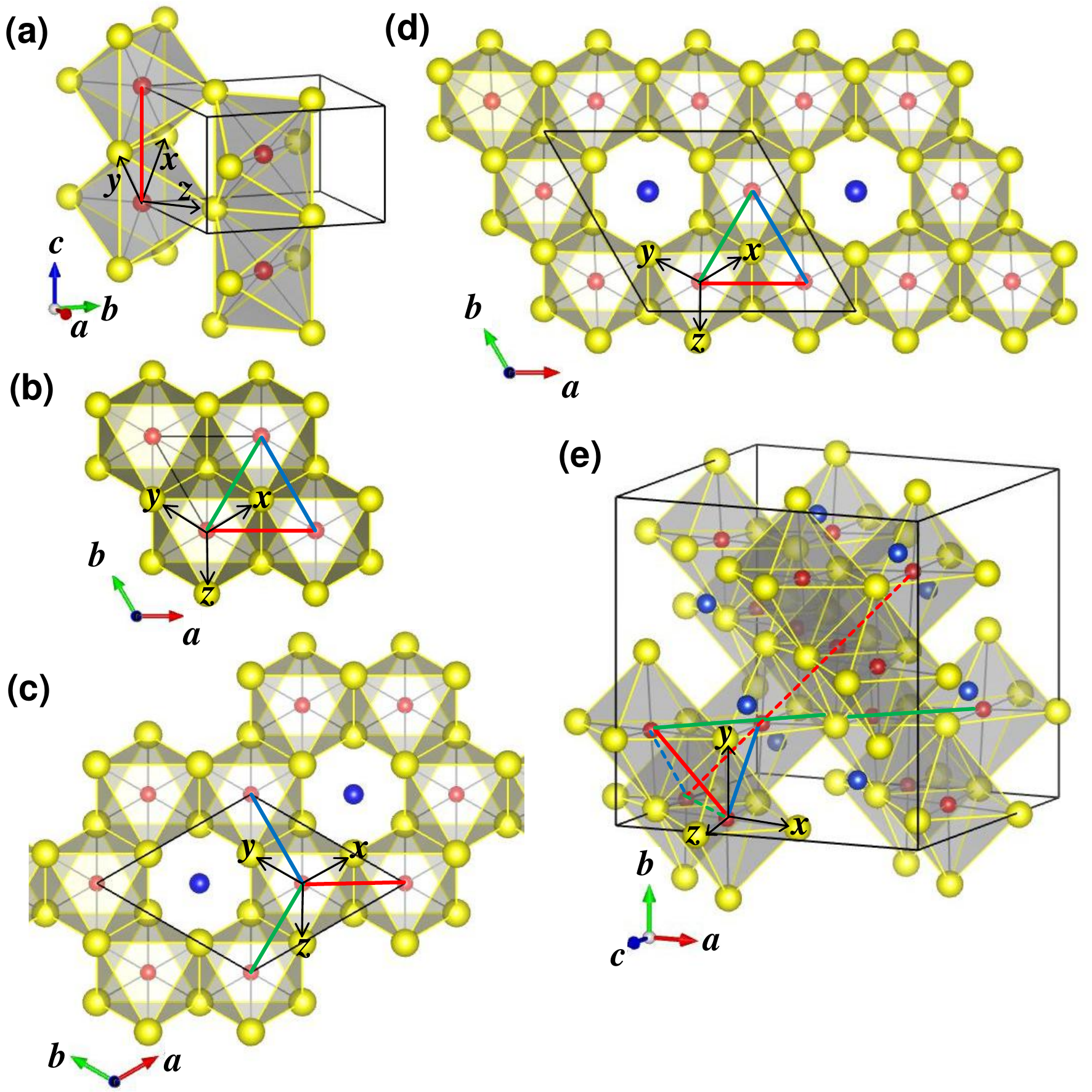}
\caption{Examples of various structures with orbital-induced Peierls transition:  (a) rutile-type structure containing chains of edge-sharing $MX_6$ octahedra, (b) CdI$_2$-type ($\alpha$-NaFeO$_2$ has very similar motif) structure having triangular lattice of transition-metal sites, (c) ilmenite-type structure (transition-metal sites form honeycomb network), (d) nolanite-type structure (transition-metal sites form kagome network), and (e) spinel-type structure (transition-metal sites form pyrochlore network consisting of metal tetrahedra). The $MX_6$ octahedra are shaded. The transition-metal $M$ and ligand $X$ are indicated by red and yellow balls. In (c) and (d), alkaline or alkaline-earth cations partly substituted for transition-metal ions of (b) are indicated by blue balls. In (e), A-site cations are indicated by blue balls. The $x$-, $y$-, $z$-axes are along the $M$-$X$ bonds. The $M$-$M$ bonds along [1,$\pm$1,0], [0,1,$\pm$1], and [$\pm$1,0,1] are indicated by red, blue, and green lines, respectively. Structures were generated using VESTA \cite{Vesta}.} 
\label{structure}
\end{figure}

There are many interesting edge-sharing transition-metal compounds with partially filled $t_{2g}$ orbitals. As described above, the direct hopping between $t_{2g}$ orbitals of the neighboring sites plays the major role in the edge-sharing systems. As typical examples, rutile-type, CdI$_2$-type, ilmenite-type, nolanite-type, and spinel-type structures are illustrated in Figs.~\ref{structure}(a), (b), (c), (d), and (e), respectively. In the rutile-type structure, the edge sharing chains run along the $c$-axis, and the $xy$ orbitals have the substantial direct hopping along it. In the CdI$_2$-type structure, the edge sharing octahedra form the triangular lattice in the $ab$-plane. The same motif of the $MX_6$ octahedra packing is observed in $\alpha$-NaFeO$_2$ structure.  The $xy$, $yz$, and $zx$ orbitals respectively have the direct hopping along the [1,-1,0], [0,1,-1], and [1,0,-1] directions. The edge sharing octahedra form the honeycomb network and kagome network in the ilmenite-type structure and the nolanite-type structure, respectively, in which part of the transition-metal $M$ ions in the triangular lattice are periodically replaced by alkaline and alkaline-earth cations as illustrated in Fig.~\ref{structure}(c) and (d).
In the spinel-type structure, the edge sharing octahedra form the pyrochlore network and the $xy$, $yz$, and $zx$ orbitals respectively have the direct hopping along the [1,$\pm$1,0], [0,1,$\pm$1], and [1,0,$\pm$1] directions.

\begin{table}[t!]
\begin{tabular}{llllll}
\hline
\hline 
Material  & Occ. &  Lattice & $T_s$ & Cluster \\
\hline

MgTi$_2$O$_4$ &  $d^{1}$ & Spinel & 260K~\cite{Isobe2002a} & dimer~\cite{Khomskii2005a}\\

NbO$_2$ &  $d^{1}$ & Rutile & 1081K~\cite{Janninck1966} & dimer~\cite{Eyert2001}\\

NaTiSi$_2$O$_6$ &  $d^1$ & Pyroxene & 210K~\cite{Isobe2002} & dimer~\cite{Wezel2006,streltsovComment2006a,Feiguin2019a}\\

NaTiO$_2$ &  $d^1$ & Triangular & 250K~\cite{Takeda1992,Clarke1998} & dimer\\

MoO$_2$ &  $d^{2}$ & Rutile & - \cite{Magneli1955} & dimer~\cite{Eyert2000}\\

Na$_2$Ti$_3$Cl$_8$ &  $d^2$ & Kagome & 200K~\cite{Hinz1995,Hanni2017a,Kelly2019a} & trimer~\cite{Khomskii2021comment}\\

LiVO$_2$ &  $d^2$ & Triangular & 500K~\cite{bongers1957structuur,Pen1997,Ponosov2024,Yun2025} & trimer~\cite{Khomskii2021a}\\

BaV$_{10}$O$_{15}$ &  $d^{2.2}$ & Triangular & 123K~\cite{Kajita2010,Takubo2012} & trimer\cite{Yoshino2017}\\

AlV$_2$O$_4$ &  $d^{2.5}$ & Spinel & 700K~\cite{Matsuno2003,Browne2017} & tri\&tetramer~\cite{Okawa2024}\\

GaV$_2$O$_4$ &  $d^{2.5}$ & Spinel & 415K~\cite{Browne2018} & tri\&tetramer\\

ReS$_2$(Se$_2$)  &  $d^3$ & Triangular & -~\cite{Murray1994,Lamfers1996} & chain~\cite{Khomskii2021a}\\

IrTe$_2$ &  $d^{5}$ & Triangular & 270K~\cite{Jobic1991,Matsumoto1999} & dimer~\cite{Ootsuki2012}\\

RuP & $d^{5}$ & MnP & 260K~\cite{hirai2012,koch2022} & trimers~\cite{hirai2022}\\

LiRh$_2$O$_4$ &  $d^{5.5}$ & Spinel & 170K~\cite{Okamoto2008,Knox2013} & dimer~\cite{Okamoto2008}\\
 
CuIr$_2$S$_4$ &  $d^{5.5}$ & Spinel & 230K~\cite{Radaelli2002,Ohashi2025} & dimers~\cite{Khomskii2005a}\\
\hline
\hline
\end{tabular}
\caption{Selected materials whose structural transitions at $T_s$ or dimerized structures are explained by the orbital-induced Peierls transition. If clusters exist at all temperatures; this is indicated by ``$-$''.
Last column contains references to theoretical studies related to this mechanism. Materials with a honeycomb lattice are not included in this list. Dimerization in this case is also governed by orbital degrees of freedom, but via local physics related to the formation of molecular orbitals rather than band effects, see also Experimental methods section. We also note that due to the large crystal-field splitting $t_{2g}$ states are filled in the $4d$ and $5d$ transition metals such as Ir, Rh, and Ru.
}
\label{Tab:Materials}
\end{table}

\hfill \break {\bf \noindent Peierls OR NOT REALLY Peierls: LOCAL EFFECTS\\}
The Peierls transition is often considered as a band phenomenon, with distortions formed spontaneously at a certain temperature $T_s$ and associated with the energy gain due to gap formation in the band spectrum. 
However, modern experimental methods such as pair distribution function (PDF) analysis of X-ray or neutron diffraction data show that dimers or other clusters typically survive even in the high-temperature phase, see Calculation and experimental methods section and, e.g.,~\cite{Bozin2019,Kojima2019,Browne2020,Koch2021,Kojima2023}, whereas a simplistic theory of the Peierls transition would not predict this. Such clusters are sometimes referred to as orbital molecules~\cite{attfield2015}.

The energy gain due to the formation of bonding states in a cluster is typically much larger than the gap in the band spectrum in a ``pure'' Peierls theory. Therefore, while the long-range order of clusters disappears at $T_s$, the clusters themselves still exist even at much higher temperatures. Such a state may be considered as a valence bond liquid with clusters floating over the lattice~\cite{Kimber2014}. In a sense, Peierls physics can be considered a source of instability or a driving force, whereas local chemical bonding in clusters is characterized by very different energetics and often cannot be broken easily.

There are also other important local effects in addition to molecular bonding. First of all, these are the correlation effects related to Coulomb interactions between electrons. In the simplest approach, molecular orbitals are linear combinations of atomic orbitals (MO LCAO). In the case of an isolated dimer with two electrons, the corresponding wavefunction is
\begin{eqnarray}
\Psi_{MO} = \frac 12 
(c^{\dagger}_{1\uparrow} +c^{\dagger}_{2\uparrow})
(c^{\dagger}_{1\downarrow} +c^{\dagger}_{2\downarrow}) | 0\rangle,
\end{eqnarray}
where $| 0\rangle$ is the vacuum state, 1 and 2 denote the two atoms, and term $c^{\dagger}_{1\uparrow} +c^{\dagger}_{2\uparrow}$ (or $c^{\dagger}_{1\downarrow} +c^{\dagger}_{2\downarrow}$) creates a molecular orbital occupied by spin up (or down). $\Psi_{MO}$ contains terms with two electrons being placed at the same site ($c^{\dagger}_{1\uparrow} c^{\dagger}_{1\downarrow}$ and $c^{\dagger}_{2\uparrow} c^{\dagger}_{2\downarrow}$). On-site repulsion of electrons destabilizes such a state and, in the limit of large Hubbard $U$, leads to the Heitler-London (HL) wavefunction
\begin{eqnarray}
\Psi_{HL} = \frac 1{\sqrt 2} 
(
c^{\dagger}_{1\uparrow}c^{\dagger}_{2\downarrow} -
c^{\dagger}_{1\downarrow}c^{\dagger}_{2\uparrow}
) 
| 0\rangle
\end{eqnarray}
while the intermediate regime can be described by Coulson-Fisher theory~\cite{Coulson1949}. Similar local physics related to electron-electron interaction is relevant for concentrated crystalline solids. 

Hubbard $U$ can localize electrons at particular sites without forming molecular orbitals of any kind. Indeed, CuO$_4$ plaquettes share common edges in the famous CuGeO$_3$, but all $t_{2g}$ orbitals of Cu$^{2+}$ are occupied. Consequently, there is no energetic gain from forming molecular orbitals, and we end up with localized electrons and $S = 1/2$ per site. Nevertheless, the corresponding chain can dimerize and gain energy by forming spin singlets ($S_{tot} = 0$ from two $S = 1/2$) via the spin-Peierls mechanism~\cite{Hase1993,Bulaevskii1978,Buzdin1980}. This state, however, is very different from both the MO LCAO and HL pictures explained above.

In the strong-coupling (large $U$) regime of materials with orbital degrees of freedom, orbital ordering often occurs~\cite{Khomskii-book1}. However, (orbital) quantum fluctuations have been theoretically proposed to stabilize exotic phases with static or dynamical spin and orbital dimerization, owing to the strong exchange interactions in the orbital sector described by corresponding pseudospins (see, e.g., \cite{Pati1998,Horsh2003}). This possibility is referred to as the orbital Peierls effect and represents an extremely interesting phenomenon. This coupling might effectively quench quantum fluctuations by locking the orbitals into static patterns, thereby rendering the orbital order largely classical in nature.

In real transition-metal compounds the situation can be even more complex because of a larger number of orbitals, spin-orbit coupling, effects of ligand crystal field, band structure due to other electrons, magnetism, etc. Therefore, even the seemingly simple situation of V$^{4+}$ ($d^1$) dimers in VO$_2$ still attracts a lot of attention.

While considerable progress had already been achieved in the early stages of Peierls transition studies by taking into account electron-phonon interaction of different forms~\cite{Bulaevskii1975}, disorder~\cite{lloyd1969,thouless1972}, and fluctuations~\cite{lee1973}, a true many-electron theory for real materials that describes both the dimerized and the high-temperature disordered phase, including all relevant local and non-local interactions, has yet to be developed.

In fact, local and band effects are hard, if possible at all, to disentangle. The instability due to nesting in the band spectrum and the tendency to form local clusters typically work hand in hand. Therefore, in this review, we do not restrict ourselves to considering only situations with a ``canonical'' Peierls transition, but also discuss materials where orbital degrees of freedom induce the formation of various clusters due to other (local) factors.

\begin{figure}
 \centering
\includegraphics[width=0.6\textwidth]{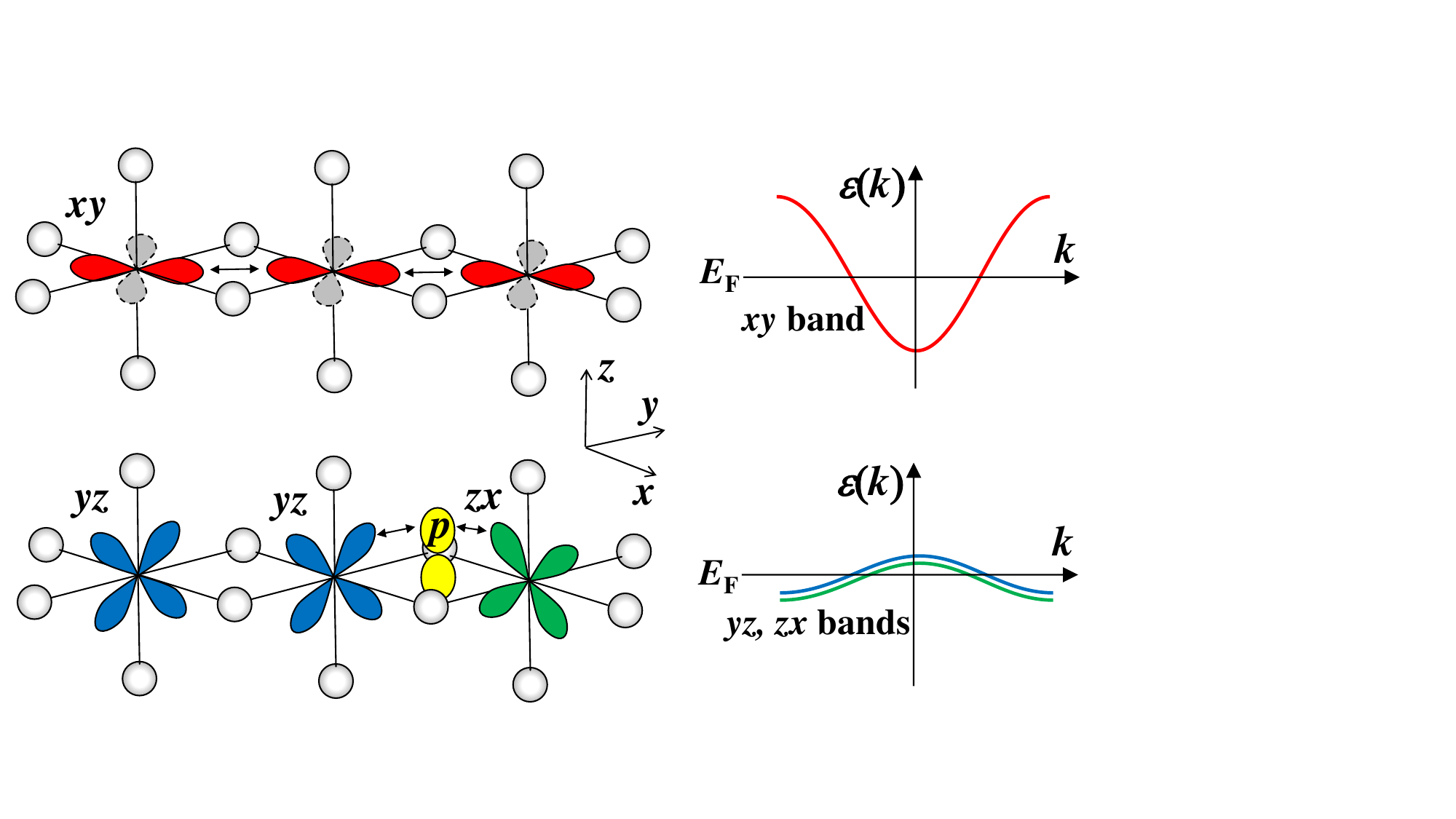}
\caption{In edge-sharing geometry, there are two types of orbitals: the $xy$ orbitals have a strong $d$-$d$ hopping and large bandwidth susceptible to the Peierls transition, while $xz$ and $yz$ orbitals form $\pi$ bonds with ligands $p$ orbitals, their bandwidths are typically smaller and they often behave as more correlated with localized electrons.} 
\label{1D-orbitals}
\end{figure}

\hfill \break {\bf \noindent QUASI ONE-DIMENSIONAL LATTICES\\}
In quasi-1D edge-sharing systems such as rutile, pyroxene, or hollandite, the $t_{2g}$ orbital degeneracy is already lifted due to the anisotropic crystal structure of the quasi 1D arrangement of the $MX_6$ octahedra. Yet, the orbital degrees of freedom can play significant roles in the rich electronic and lattice transitions in these systems. Indeed, in this geometry one of the $t_{2g}$ orbitals provides a direct $d$-$d$ overlap and form a wide band, while others mostly participate in ligand-assisted hopping as schematically shown in Fig.~\ref{1D-orbitals}.

In the classical metal-insulator transition in rutile-type VO$_2$ (V$^{4+}$, $3d^{1}$)  \cite{Morin1959}, the $d^1$-$d^1$ dimerization is assisted by orbital reconfiguration \cite{Haverkort2005} and electronic correlation effects \cite{Biermann2005}, which can be controlled in thin films through the substrate effect \cite{Aetukuri2013}. Under the orbital-assisted mechanism, one of the $t_{2g}$ bands accommodates most of the $d$ electrons, forming a half-filled quasi-1D band.
The Fermi surface of the half-filled band tends to satisfy the nesting condition~\cite{kim2013} and the electron-phonon interaction induces the doubling of the unit cell due to dimerization along the edge-sharing chain.

The dimerization in NbO$_2$ (Nb$^{4+}$, $4d^1$) \cite{Janninck1966} having the same rutile crystal structure is attracting renewed interest due to its chaotic breakdown which can be applied to neural computing \cite{Kumar2017}. While the correlation effects are generally expected to weaken for more extended $4d$ and $5d$ wavefunctions~\cite{moore2024}, even in NbO$_2$ spectroscopic experiments \cite{OHara2014} and theoretical analysis \cite{Craco2024} suggest the importance of electronic correlations.

One of the most interesting question is how sensitive these transitions to the $d$-band filling. From the perspective of the local physics discussed in the previous section, the $d^1$-$d^1$ dimerization is expected to remain robust against hole doping. Indeed, if filling of the quasi one-dimensional band is less than 1/2, the $d^0$ sites are created and the remaining $d^1$ sites form the $d^1$-$d^1$ dimers as realized in Ti$_4$O$_7$ (Ti$^{3.5+}$, formally $3d^{0.5}$) \cite{Lakkis1974,Leonov2006}.
When the $d^1$-$d^1$ dimers are embedded between the $d^0$ sites, the HL-like singlet bond character would be enhanced \cite{Taguchi2010}.
In opposite situation of electron doping, the $d^2$ sites are created and the remaining $d^1$ sites form the $d^1$-$d^1$ dimers. Even in the $d^2$ systems such as rutile-type MoO$_2$, dimerization is realized \cite{Rogers1969,Eyert2000}, while the importance of correlation effects is suggested by photoemission spectroscopy \cite{Stoeberl2017}. This is achieved via an orbital-selective mechanism (described at the very end of this review), in which some electrons form molecular orbitals, while others remain site-localized (forming weak $\pi$ bonds)~\cite{Streltsov2014a}. On the other hand, the $d^2$-$d^2$ dimer in V$_4$O$_7$ (V$^{3.5+}$, formally $3d^{1.5}$) \cite{Hodeau1978}, the V$^{3+}$ ions with $S=1$ are antiferromagnetically coupled via the $d$-$p$-$d$ superexchange path \cite{Botana2011}.

Finally, it should be mentioned that dimerization is possible not only in edge-sharing chains discussed so far, but also in face-sharing geometries. The band which is expected to be susceptible to the Peierls instability has the $a_{1g}$ character, see Fig.~\ref{Fig:sharing}(c). Representative examples of such materials are trihalides such as TiI$_3$ (Ti$^{3+}$, $3d^1$) with $T_s = 323$~K~\cite{Angelkort2009}, as well as $\beta$-RuCl$_3$ and RuBr$_3$ (Ru$^{3+}$, $4d^5$) with $T_s = 206$~K and 384~K, respectively~\cite{Hillebrecht2004}. However, there are indications that the high-temperature structure in the latter material may only appear uniform, while in fact being a disordered version of dimerized chains due to stacking faults~\cite{Merlino2004}. This again points to the importance of local effects, such as chemical bonding.

\hfill \break {\bf \noindent TWO DIMENSIONAL LATTICES}
\hfill \break {\bf \noindent Triangular network.}
In contrast to the quasi-1D lattices, only edge-sharing is possible in 2D lattices. Most of the materials with a triangular lattice discussed below have the CdI$_2$ or $\alpha$-NaFeO$_2$ structure shown in Fig.~\ref{structure}(b).

Let us consider a $yz$/$zx$/$xy$ three-band model on a triangular lattice. Since ligand orbitals are not included in this simple model, the conventional (local) Jahn-Teller effect - which relies on distortion of the ligand octahedra to lift the degeneracy of the $t_{2g}$ subshell - is not operative in this situation. Instead, interplay between band Jahn-Teller effect and Peierls instability can be discussed. 

The hopping integrals between the $xy$, $yz$, and $zx$ orbitals along [1,-1,0], [0,1,-1], and [-1,0,1], respectively, are given by $t_{xy}$, $t_{yz}$, and $t_{zx}$ ($t_{xy}=t_{yz}=t_{zx}$ for the undistorted triangular lattice).  The $xy$, $zx$, and $yz$ orbitals form 1D bands: $\varepsilon_1 =2t_{xy}\cos(k_X)$, $\varepsilon_2 =2t_{zx}\cos(k_X/2+\sqrt{3}k_Y/2)$, and $\varepsilon_3 =2t_{yz}\cos(-k_X/2+\sqrt{3}k_Y/2)$, respectively. Here, $k_X$ and $k_Y$ are the wave numbers along the [1,-1,0] and [1,1,-2] directions in the unit of $1/a$ ($a$ is the lattice constant of the triangular lattice). 

\begin{figure}[b!]
 \centering
\includegraphics[width=0.45\textwidth]{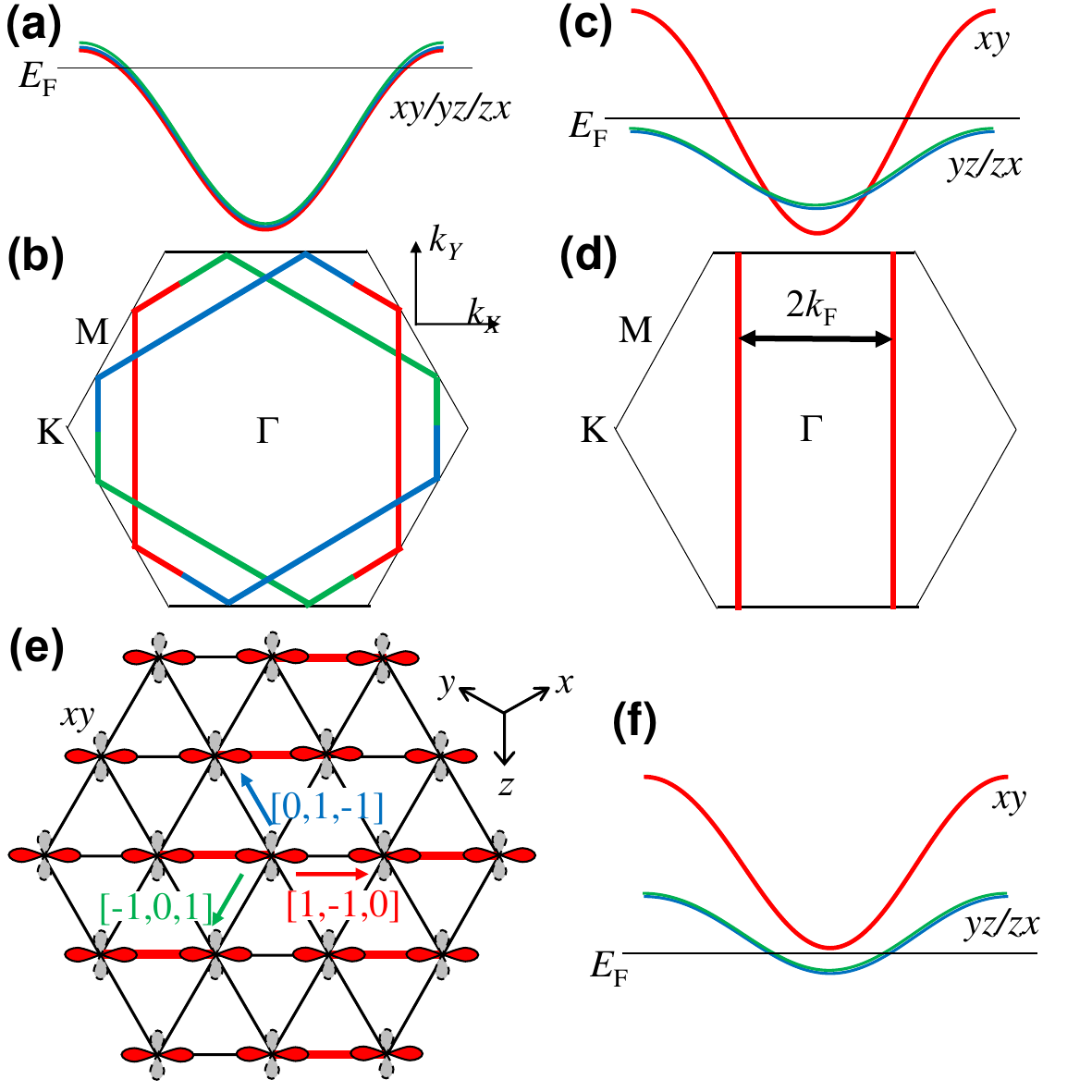}
\caption{(a) Band dispersion for an undistorted triangular lattice three-band model ($yz$/$zx$/$xy$ orbitals) with one $t_{2g}$ hole per site. The $t_{2g}$ hole is shared by the $xy$, $yz$, and $zx$ bands. 
(b) Fermi surfaces for the undistorted triangular lattice three-band model with one $t_{2g}$ hole per site. 
(c) Band dispersion for a distorted triangular lattice (compressed along [1,-1,0] direction). The $t_{2g}$ hole is accommodated by the $xy$ band.  
(d) Fermi surfaces for the distorted case. (e) Orbital ordering and dimerization by the orbital-induced Peierls effect on the triangular lattice. Short metal-metal bonds and active lobes of the $xy$ orbitals are shown in red. (f) Possible band dispersion for a distorted triangular lattice with one $t_{2g}$ electron per site.} 
\label{tri}
\end{figure}

As an example, let us consider the three one-dimensional bands  with a single hole, as shown in Fig.~\ref{tri}(a). These give rise to one-dimensional Fermi surfaces with 5/6 filling, as illustrated in Fig.~\ref{tri}(b). In real materials, there are also ligand-assisted hoppings ($t_{yz-zx}$, $t_{zx-xy}$, and $t_{xy-yz}$) mixing these bands, which can lead to imperfect Fermi surface nesting. 

The degeneracy of the $xy$, $yz$, and $zx$ bands can be lifted by band Jahn-Teller type compression along [1,-1,0]. Then $t_{xy} > t_{yz} = t_{zx}$ and only the wider $xy$ band may accommodate the $t_{2g}$ hole as shown in Fig.~\ref{tri}(c). The $xy$ band forms a 1D band with nesting vector 2$k_{\rm F}$ as illustrated in Fig.~\ref{tri}(d). The orbital-induced Peierls state is robust against the indirect transfer terms $t_{xy-yz}$ and \textcolor{magenta}{$t_{xy-zx}$} due to the Jahn-Teller energy splitting between the $xy$ and $yz$/$zx$ orbitals. In the presence of  of $t_{xy-yz}$ and \textcolor{magenta}{$t_{xy-zx}$}, a quasi-1D Fermi surface dominated by the $xy$ character still keeps the nesting condition. Since the $xy$ band accommodates one electron and is half-filled, the periodicity along the [1,-1,0] direction is doubled with $xy$-$xy$ dimers as shown in Fig.~\ref{tri}(e).

In the orbital-induced Peierls mechanism, the $t_{2g}$ orbital degeneracy is broken by either a local Jahn-Teller effect (which splits localized levels) or a band Jahn-Teller effect (which not only splits bands but also alters their widths - a factor that can be even more important). This leads to an instability, and the resulting quasi-1D electronic state is further stabilized by a Peierls transition. The band Jahn-Teller effect would be relevant for IrTe$_2$ (one hole in $t_{2g}$) with relatively large width of Ir 5$d$ band.
On the other hand, in NaTiO$_2$ with smaller width of Ti 3$d$ band, the local Jahn-Teller distortion may play a role as discussed below. In order to describe NaTiO$_2$, the O 2$p$ orbitals should be explicitly considered. In the case of one $t_{2g}$ electron per site, the band Jahn-Teller effect may not work and modification of the bandwidth may be \textcolor{magenta}{not} so critical. Whereas the $xy$ bandwidth is increased, the center of the $xy$ band is shifted to the higher energy due to a local Jahn-Teller effect as illustrated in Fig.~\ref{tri}(f).

Among the triangular lattice systems, NaTiO$_2$ (Ti$^{3+}$, $3d^1$) exhibits a transition to the nonmagnetic insulating state with lattice distortion from trigonal to monoclinic \cite{Takeda1992,Clarke1998}.
The $xy$, $yz$, and $zx$ orbitals can form 1D bands with the direct $t_{2g}$-$t_{2g}$ transfer term, which may induce band Jahn-Teller distortion and subsequent Peierls instability. 
In these layered structures a trigonal field with the trigonal axis pointing perpendicular to the layers is often very important. It breaks the three-fold degeneracy and the $t_{2g}$ orbitals are split into the $a_{1g}$ [$\frac{1}{\sqrt{3}}(xy + yz + zx)$] and $e_{g}^{\pi}$ [$\frac{1}{\sqrt{3}}(xy+e^{\pm2\pi \rm{i}/3}yz+e^{\pm4\pi \rm{i}/3}zx)$] orbitals. 
As a result, the strong mixing between the $xy$, $yz$, and $zx$ orbitals provide the one hole pocket at the zone center (with $a_{1g}$ character) and the six hole pockets around the M points (with $e_{g}^{\pi}$ character) for the $d^1$ system. The Fermi surfaces calculated for the trigonal phase \cite{Subedi2017} are roughly consistent with this simple picture except the hole pocket at the zone center. The absence of the hole pocket can be assigned to the trigonal ligand field which stabilizes the $a_{1g}$ orbital. 

\begin{figure}[t!]
 \centering
\includegraphics[width=0.7\textwidth]{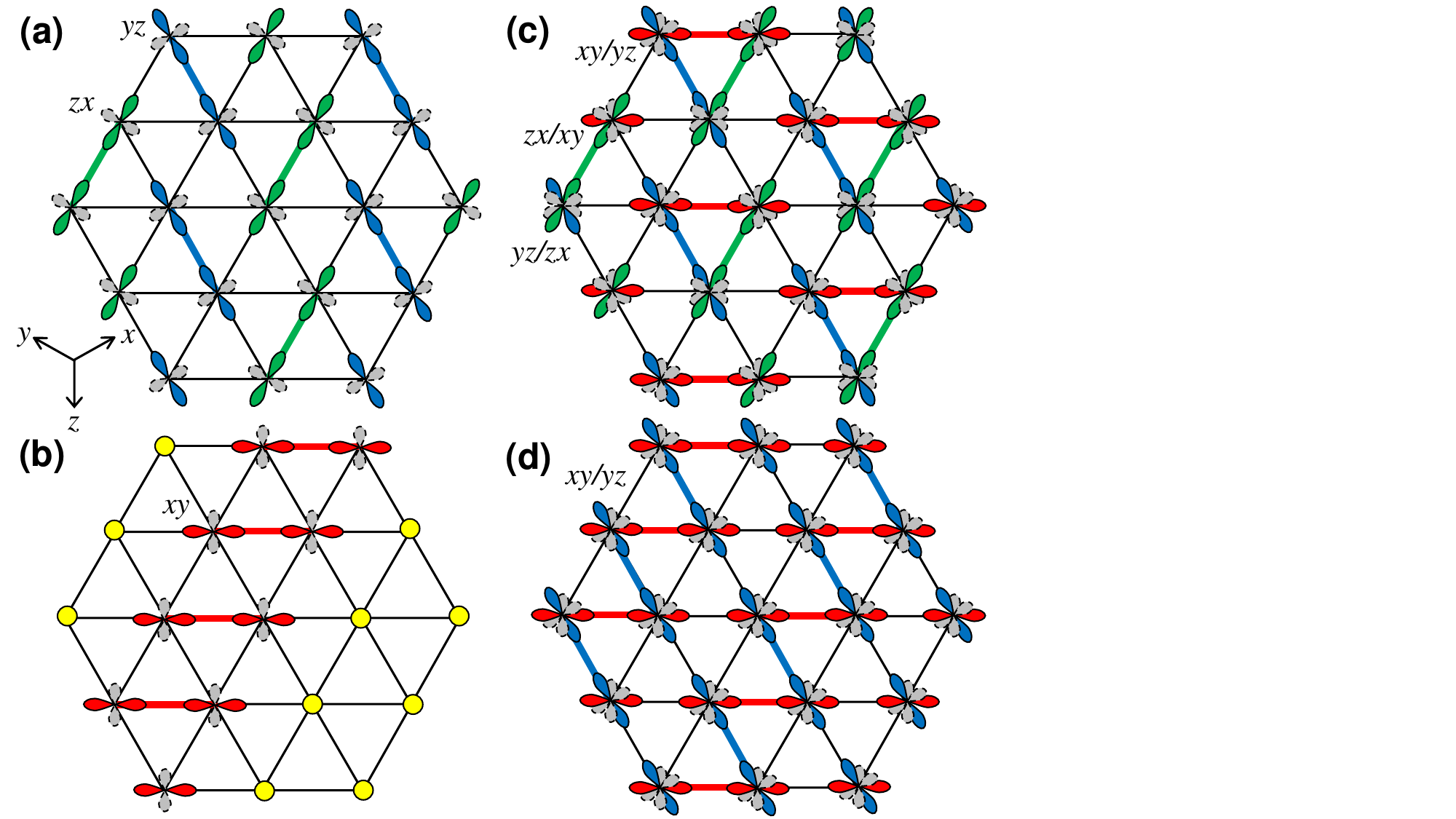}
\caption{(a) Possible orbital ordering in NaTiO$_2$. (b) Stripe-type charge and orbital ordering in IrTe$_2$; Ir ions with nearly completely filled $t_{2g}$ shell, i.e. close to 3+ are shown by yellow balls. (c) Orbital ordering and trimers in LiVO$_2$ and LiVS$_2$. (d) Orbital ordering and zig-zag bonds in LiVS$_2$. The thick solid lines indicate the short bonds of dimers and trimers. Active lobes of the $t_{2g}$ orbitals (i.e., those involved in strong bonding in clusters or the Peierls instability) are shown in red, blue, and green.}
\label{tri2}
\end{figure}

Although the band structure of the undistorted state is affected by the trigonal ligand field, the synergetic effect of the band Jahn-Teller distortion and the Peierls distortion may reorganize the band structure. Contraction along the [1,-1,0] direction makes the $xy$ band wider and also pushes it up in energy. Then, a single electron can be accommodated in the quasi 1D $xz/yz$ bands as shown in Fig.~\ref{tri}(f), they become 1/4-filled and this results in dimerization with four times periodicity as shown in Fig.~\ref{tri2}(a). The orbital polarization of NaTiO$_2$ is likely to be induced by the local Jahn-Teller effect rather than the band Jahn-Teller one. However, it is still awaiting experimental and theoretical studies to resolve the inconsistency between the lattice distortion and the spin singlet state.

It is interesting to mention a NaTiO$_2$ analogue on a square lattice (in fact, an inverse K$_2$NiF$_4$ structure): Na$_2$Ti$_2$Sb$_2$O, in which the same Ti$^{3+}$ ions with $3d^1$ configuration are separated by oxygen (i.e. through common-corner sharing). This prevents dimerization (negatively charged ion in between), but leads to a charge-density wave with suppressed, though not-zero, magnetic susceptibility~\cite{ozawa2001,davies2016,ren2017} similar to many Peierls materials. 

IrTe$_2$ (formally Ir$^{4+}$, $5d^5$) with the Ir triangular lattice exhibits a structural transition at $\sim$ 270 K from the trigonal to the monoclinic phase accompanied by anomalies in electrical resistivity and magnetic susceptibility \cite{Jobic1991,Matsumoto1999,Pyon2012,Yang2012}. An electron diffraction study shows that the structural transition is accompanied by a superstructure with wave vector of $\vec{q}$ = (1/5,0,-1/5), (1/6,0,-1/6) and (1/8,0,-1/8)  \cite{Yang2012,Pascut2014,Toriyama2014,Takubo2018}. Such superstructure can be explained by CDW driven by a perfect or partial nesting of multi-band Fermi surfaces \cite{Yang2012}. In addition, the monoclinic distortion suggests a band Jahn-Teller-like effect and the orbital-induced Peierls instability, which is partly supported by photoemission measurements  \cite{Ootsuki2012,Ootsuki2013}. The periodicity of the superstructure is not consistent with the simple dimer model shown in Fig.~\ref{tri}(e). Te has a lower electronegativity than oxygen (as in NaTiO$_2$); it cannot fully oxidize Ir to Ir$^{4+}$ (which would require removing four electrons), so IrTe$_2$ remains metallic with Ir having a mixed valence (3+/4+) \cite{Oh2013}. Structurally, this results in charge stripes consisting of Ir$^{4+}$-Ir$^{4+}$ dimers and Ir$^{3+}$ sites~\cite{Pascut2014,Toriyama2014}, as illustrated in Fig.~\ref{tri2}(b), while a particular ordering or dimers is controlled by elastic forces~\cite{Artyukhin2024}.
Domain structure \cite{Oike2018,Mizokawa2022,Nicholson2024} and the photoinduced effect \cite{Ideta2018} of the anisotropic arrangement of the dimers have been studied by means of various spectromicroscopy techniques.

LiVO$_2$ is a classical system with V$^{3+}$ ($3d^2$) which exhibits a magnetic-nonmagnetic transition around 500 K with V trimerization \cite{Rudorff1954,Tian2004,Jinno2013,Kojima2019}. Pen et al. pointed out that two of the $xy$, $yz$, and $zx$ orbitals are occupied at the V sites connected by the [1,-1,0], [0,1,-1], and [-1,0,1] bonds of the triangular lattice forming the spin singlet state \cite{Pen1997}. In the limit of weak Hund's coupling, the $d^2$ trimers on the triangular lattice correspond to a kind of orbital-induced Peierls state without global orbital polarization (while at each V particular orbitals are occupied as discussed below). For the case of $d^1$ discussed in the previous paragraphs, one of the $t_{2g}$ orbitals should be selected for the subsequent Peierls instability. In the $d^2$ configuration of LiVO$_2$, at each V site, two $t_{2g}$ electrons are distributed to three chains running through it. The $xy$, $yz$, and $zx$ bands along the three chains running along the [1,-1,0], [0,1,-1], and [-1,0,1] directions are 1/3-filled and can induce orbital ordering and dimerization with three times periodicity.  

Under the dimerization along the three directions, trimers are formed with $yz$/$zx$, $zx$/$xy$, and $xy$/$yz$ orbitals occupied at the three V sites, resulting in a nonmagnetic ground state as shown in Fig.~\ref{tri2}(c). Although the multiplet structure of V 2$p$ X-ray absorption spectroscopy of LiVO$_2$ indicates localized V 3$d$ electrons with substantial Hund's coupling \cite{Pen1997b}, such a simple band picture is still useful to understand the nonmagnetic ground state with the trimers. An alternative would be the formation of total $S_{\rm tot}=0$ from three $S=1$ spins, but the transition temperature $T_s = 500$~K appears suspiciously high for this scenario.

LiVS$_2$ (the same V$^{3+}$, $t^2_{2g}$) exhibits a metal-insulator transition at 314 K which is accompanied by V trimerization similar to LiVO$_2$ \cite{Katayama2009}. Since the V 3$d$ electrons are more itinerant in LiVS$_2$ than in LiVO$_2$, the effect of Hund's coupling is expected to be weaker. Since ratio of the band gap, $E_G$, to transition temperature $T_s$ is large $E_G/k_BT_c$ = 6 \cite{Tanaka2009}, the short range order of the short V-V bonds may survive even above $T_s$ and, indeed, the zigzag chain correlations are observed in this phase \cite{Katayama2021}. In contrast to the trimer state in Fig.~\ref{tri2}(c), the zigzag chain structure can be stabilized by the ferro-type orbital order as shown in Fig.~\ref{tri2}(d). For example, when the $xy$ and $zx$ orbitals are occupied at every V site, the $xy$-$xy$ and $zx$-$zx$ dimers form the zigzag chains as indicated by the thick lines in the figure.

Li$_2$MoO$_3$ (Mo$^{4+}$,  $4d^2$) in contrast to many other Li$_2$MO$_3$ materials discussed in the next subsection has $\alpha-$NaFeO$_2$ structure with triangular lattice. It exhibits Mo trimers similar to LiVO$_2$\cite{James1988}, which can be randomly mixed with Li ions~\cite{savina2022}. The Mo trimers can survive upto the highest temperature. Recent electrochemical study combined with DFT calculations show that molecular bonding is rather robust and suggest formation of Mo-Mo dimers under delithiation~\cite{savina2022}. 

BaV$_{10}$O$_{15}$ (V$^{2.8+}$, $3d^{2.2}$) with V$^{2+}$/V$^{3+}$ mixed valence exhibits a structural transition at 123 K driven by V trimerization (small triangles) \cite{Kajita2010} and V 3$d$ orbital order similar to LiVO$_2$ \cite{Takubo2012}. BaV$_{10}$O$_{15}$ also has a triangular-like lattice, but with part of the V sites removed and adjacent V layers attached to each other (so that the inter-layer coupling is important \cite{Yoshino2017}). Having extra $d$ electrons than in LiVO$_2$ one can occupy other orbitals. It is possible to create molecular orbitals from them and to put an extra electron in the bonding molecular orbital. Therefore, the trimer is expected to be stable against electron doping (up to the doping level of 1/3 per V site). Indeed, the V trimer in BaV$_{10}$O$_{15}$ is likely to accommodate one extra electron. The extra electron would be shared by the three V sites or localized on one of the V sites. The V$^{2+}$/V$^{3+}$ charge disproportionation is observed by X-ray photoemission spectroscopy \cite{Yoshino2017}, indicating that the ground state of the trimer would be a superposition of V$^{2+}$-V$^{3+}$-V$^{3+}$, V$^{3+}$-V$^{2+}$-V$^{3+}$, and V$^{3+}$-V$^{3+}$-V$^{2+}$ configurations. However, since one of the VO$_6$ octahedron in the trimer shares its face with the one in the neighboring layer, the extra electron would be localized at this V site \cite{Yoshino2017}.

\begin{figure}[t!]
 \centering
\includegraphics[width=0.4\textwidth]{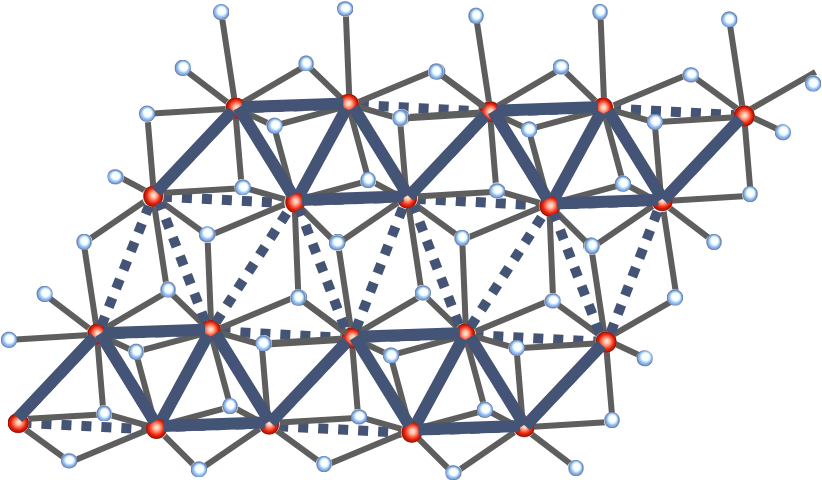}
\caption{Crystal structure of $1T$ polymorphs of ReS$_2$, ReSSe, ReSe$_2$, TcS$_2$, as well as LiMoS$_2$ and NaMoO$_3$, where dimerization in three different directions results in a ``diamond necklace'' pattern. Metal are shown as red balls, ligands as white balls, and metal-metal dimers as bold dark blue lines.}
\label{ReS2}
\end{figure}

Even more exotic dimerization pattern is observed in the $1T$ polymorphs of ReS$_2$, ReSSe, and ReSe$_2$ (Re$^{4+}$, $5d^3$)~\cite{Murray1994,Lamfers1996}. In this case, we have one electron in each $t_{2g}$ band (i.e., half-filling) and neglecting correlation effects, Hund's exchange and spin-coupling one can expect dimerization in all three directions, exactly as in experiment, see Fig.~\ref{ReS2}. These dimers are stacked in the ``diamond necklace'' structure, which, as we see, can be explained by the orbital-induced Peierls mechanism~\cite{Khomskii2021a}. Moreover, TcS$_2$ (Tc$^{4+}$, $4d^3$)~\cite{Lamfers1996}, LiMoS$_2$ (Mo$^{3+}$, $4d^3$)~\cite{schwarzmuller2024}, and even NaMoO$_3$ (Mo$^{3+}$, $4d^3$)~\cite{vitoux2020}, which has a very different $\alpha$-NaFeO$_3$ crystal structure, show the same pattern. It should be noted that the orbital-Peierls-like treatment can predict periodicity, but it is typically unable to lock the phase between the chains (i.e., how they are stacked in the lattice). This last problem can be solved, for example, by considering extensions of loop models~\cite{knowles2025}.

\begin{figure}[t!]
 \centering
\includegraphics[width=0.79\textwidth]{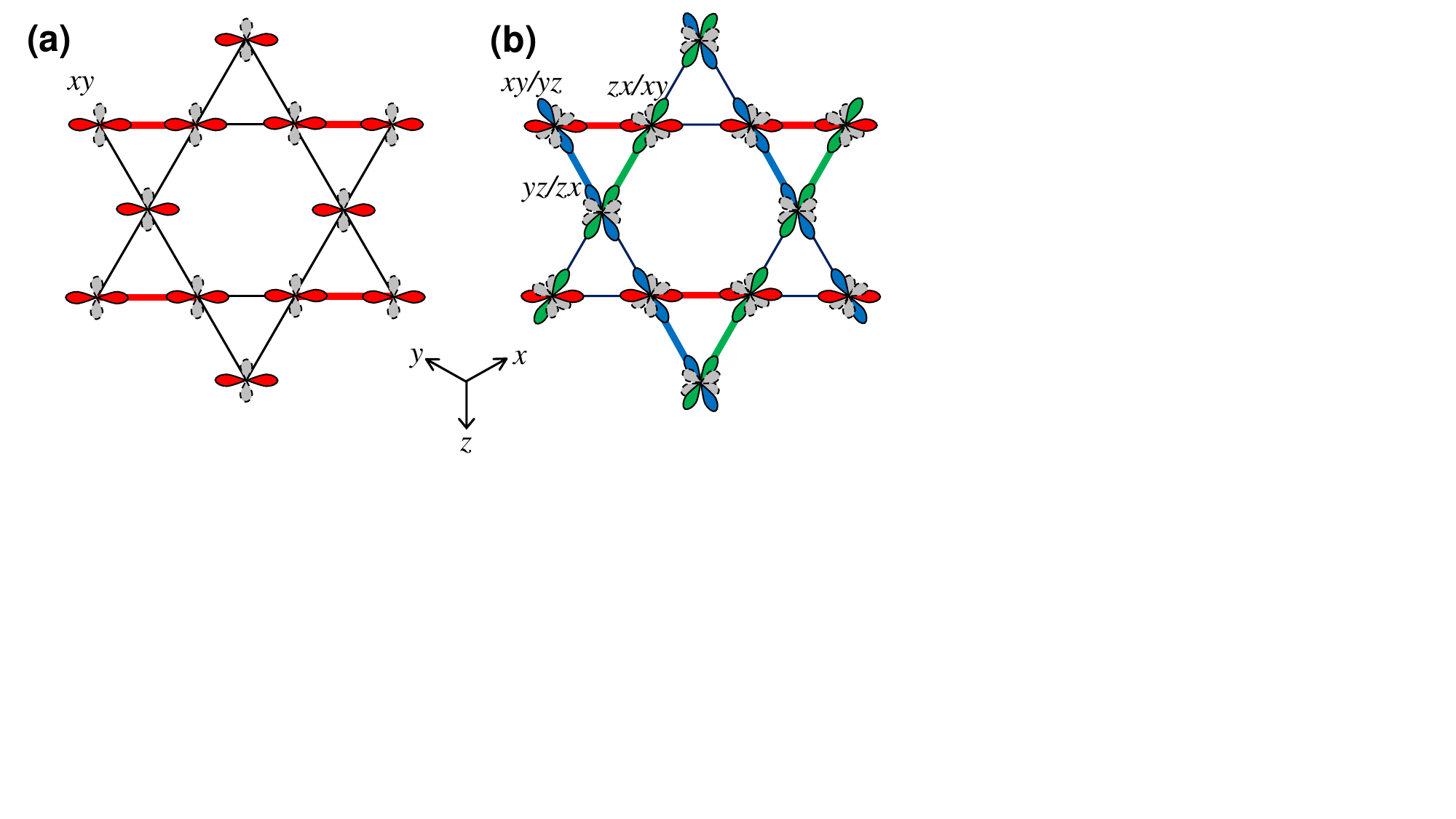}
\vspace{-3.5cm}
\caption{Kagome network. (a) Orbital ordering for $d^1$ on a kagome network. (b) Orbital ordering for $d^2$ on a kagome network in Na$_2$Ti$_3$Cl$_8$. The same trimerization is observed in Zn$_2$Mo$_3$O$_8$, LiZn$_2$Mo$_3$O$_8$ and many other Mo$_3$O$_8$-based materials. The thick solid lines indicate the short bonds. Active lobes of the $t_{2g}$ orbitals are shown in red, blue, and green.}
\label{kagome}
\end{figure}

\hfill \break {\bf \noindent Kagome network.}
This lattice is obtained from triangular network by removing 1/6 of the sites, Fig.~\ref{kagome}. In materials considered below, transition metals are again in the octahedral surrounding and these octahedra share their edges.

In the case of $d^1$ or $d^5$ metals one of the $t_{2g}$ orbitals should be selected for the subsequent Peierls instability as shown in Fig.~\ref{kagome}(a). Such dimerized state in the kagome network cannot be nonmagnetic since 1/3 of the transition-metal sites are not involved in the dimers. On the other hand, $d^2$ or $d^4$ kagome materials are expected to have a stable nonmagnetic state.

Na$_2$Ti$_3$Cl$_8$ consists of a Ti$^{2+}$ ($3d^2$, two electrons in $t_{2g}$) kagome network and exhibits partial trimerization around 210 K and full trimerization at 190 K \cite{Hinz1995,Hanni2017a, Kelly2019a}, which leads to formation of so-called breathing kagome lattice. It is possible to describe the spin singlet trimer based on the antiferromagnetically coupled localized spin $S=1$ \cite{Paul2020}. However, the spin-lattice coupling may not be strong enough to explain the substantial lattice displacement of the trimerization and rather high temperatures structural transitions. The large lattice distortion suggests that the strong electron-lattice rather than the spin-lattice interaction plays a vital role. Therefore, although Na$_2$Ti$_3$Cl$_8$ is highly insulating, it is useful to describe the trimerization based on the itinerant picture of orbital-induced Peierls mechanism \cite{Khomskii2021comment}. Similar to the triangular lattice, the kagome network can be decomposed into three chains running along the [1,-1,0], [0,1,-1], and [-1,0,1] directions. Therefore, the $xy$, $yz$, and $zx$ orbitals can form quasi 1D bands along the three directions. At each Ti site, two $t_{2g}$ electrons are distributed to two chains running through it. Therefore, as illustrated in Fig.~\ref{kagome}(b), each 1D band is half-filled and induces the lattice modulation with two times periodicity. The short Ti-Ti bonds along the three directions form the trimers in agreement with the experimental observation.

\begin{figure}[t!]
\centering
\includegraphics[width=0.55\textwidth]{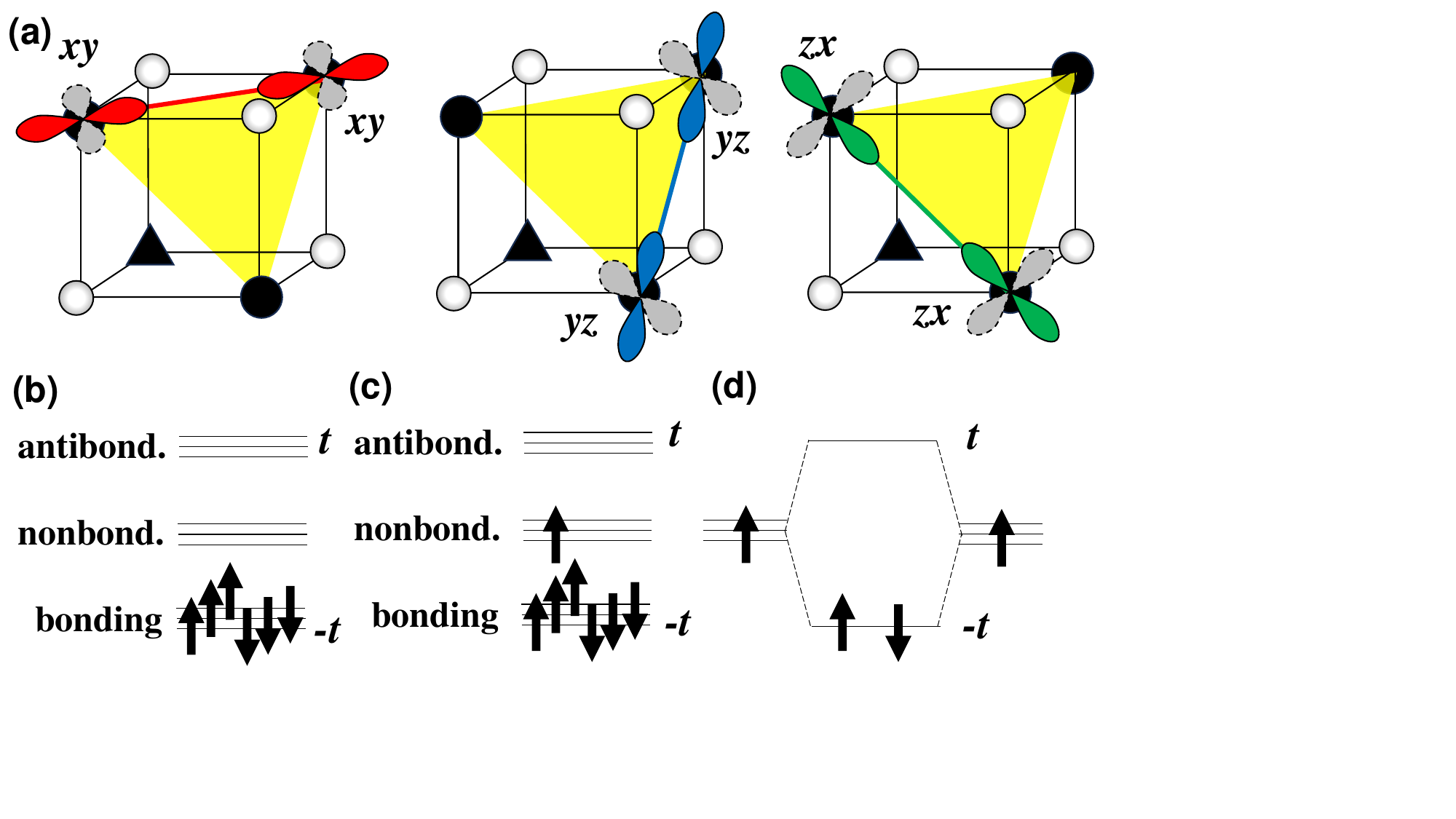}
\caption{(a) Formation of molecular bonds in trimers (shadowed triangle), where the $MX_6$ octahedra (transition metals, $M$, are shown by black balls, while ligands, $X$, by white ones) share their edges in the triangular network systems such as LiVO$_2$ and the kagome network systems such as Zn$_2$Mo$_3$O$_8$. (b) Electronic structure, when each of metal has $d^2$ electronic configuration as e.g. in Zn$_2$Mo$_3$O$_8$. (c) Situation of $d^{7/3}$ configuration as realized, e.g., in LiZn$_2$Mo$_3$O$_8$ and Nb$_3$Cl$_8$. (d) Electronic structure of Mo-Mo dimer in MoO$_2$: part of electrons occupy molecular orbitals, while the other one stay at localized atomic-like orbitals.}
\label{Fig:orbital-selectivity}
\end{figure}

The Mo$^{4+}$ ($4d^2$) kagome network in Zn$_2$Mo$_3$O$_8$ hosts Mo$_3$ trimers similar to Na$_2$Ti$_3$Cl$_8$ \cite{Sheckelton2012,Mourigal2014}. The Mo trimers with six $t_{2g}$ electrons are stable up to the highest temperature available indicating that the trimer is more stable in the more itinerant system. Moreover, they remain stable even under doping. This suggests the importance of local physics: the formation of chemical bonding. In this picture, three $t_{2g}$ orbitals form three molecular orbitals, which accommodate six electrons per trimer (see Fig.~\ref{Fig:orbital-selectivity}, which also applies to LiVO$_2$), yielding a substantial energy gain. Electron doping does little harm, as extra electrons occupy non-bonding levels, resulting in the orbital-selective behavior discussed in the last section.

Indeed, one can easily intercalate different ions between the Mo$_3$O$_8$ layers, and the materials obtained in this way exhibit extraordinary physical properties. For example, Mo$_3$ trimers survive in LiZn$_2$Mo$_3$O$_8$, where there are seven Mo $4d$ electrons per trimer~\cite{Sheckelton2012,Mourigal2014}. 
E.g. at the Mo site with $yz$ and $zx$ electrons occupied, the $xy$ orbital is available. In the three Mo sites in the timer, such “free” orbitals ($xy$, $yz$, and $zx$ at the three Mo sites) can form the “non-bonding” molecular orbital. Such “non-bonding” molecular orbital is occupied by the extra electron as illustrated in Fig.~\ref{Fig:orbital-selectivity}(b) and (c).
The Mo$_3$ trimers have localized spin 1/2 due to the extra electron and form a geometrically frustrated spin 1/2 triangular lattice. This can be viewed as a cluster Mott insulating state \cite{Chen2016,Nikolaev2021}, providing exotic spin-liquid behavior \cite{Sheckelton2012,Mourigal2014}. Mo$^{3+}$/Mo$^{4+}$ charge fluctuation is observed by X-ray photoemission spectroscopy \cite{Streltsov2025}. This situation is similar to the V$_3$ trimer in BaV$_{10}$O$_{15}$.
Possible relationship between the charge fluctuation and the spin-liquid behavior should be examined by further studies.

There is also a very large class of materials with similar breathing kagome lattices and general formula Nb$_3$X$_8$, where $X=$F, Cl, Br, and I.  Nb$_3$Cl$_8$ is one of the most studied among them. Since Nb$^{2+}$ : Nb$^{3+}$ = 1 : 2, the Nb trimer also accommodates seven $t_{2g}$ electrons and hosts localized spin 1/2 \cite{Cotton1987,Strobele2001, Haraguchi2017}. The spin-1/2 triangular lattice of the cluster Mott insulating state is a new playground to study the spin liquid and exotic superconductivity \cite{Zhang2023}. Interestingly, Nb$_3$Cl$_8$ and  Nb$_3$Br$_8$ exhibit a magnetic-nonmagnetic transition around 90 K~\cite{Sheckelton2017} and 390 K\cite{Pasco2019}, respectively, although orbital degeneracy is already lifted. Formation of a nonmagnetic state often implies dimerization, but in this case this must be dimers made out of $S_{tot}=1/2$ trimers. Different scenarios have been proposed including intra- and inter-layer dimerization\cite{Haraguchi2017,Sheckelton2017} and X-ray photoemission spectroscopy indeed observes Nb$^{2+}$/Nb$^{3+}$ charge fluctuations \cite{Nakamura2024}. Theoretical studies suggest that there is an extremely strong exchange coupling between layers, much larger than intralayer exchange interaction and transition to a nonmagnetic state can be considered as formation of dimerized bilayers~\cite{Aretz2025}.

\hfill \break {\bf \noindent Honeycomb network.}
It is not straightforward to apply the orbital-induced Peierls description to the honeycomb network since it cannot be decomposed into chains. This suggests importance of local physics related to formation of clusters due to chemical bonding in which orbital degrees of freedom play a very important role (see also the Peierls or not really Peierls section above).

\begin{figure}[t!]
 \centering
\includegraphics[width=0.7\textwidth]{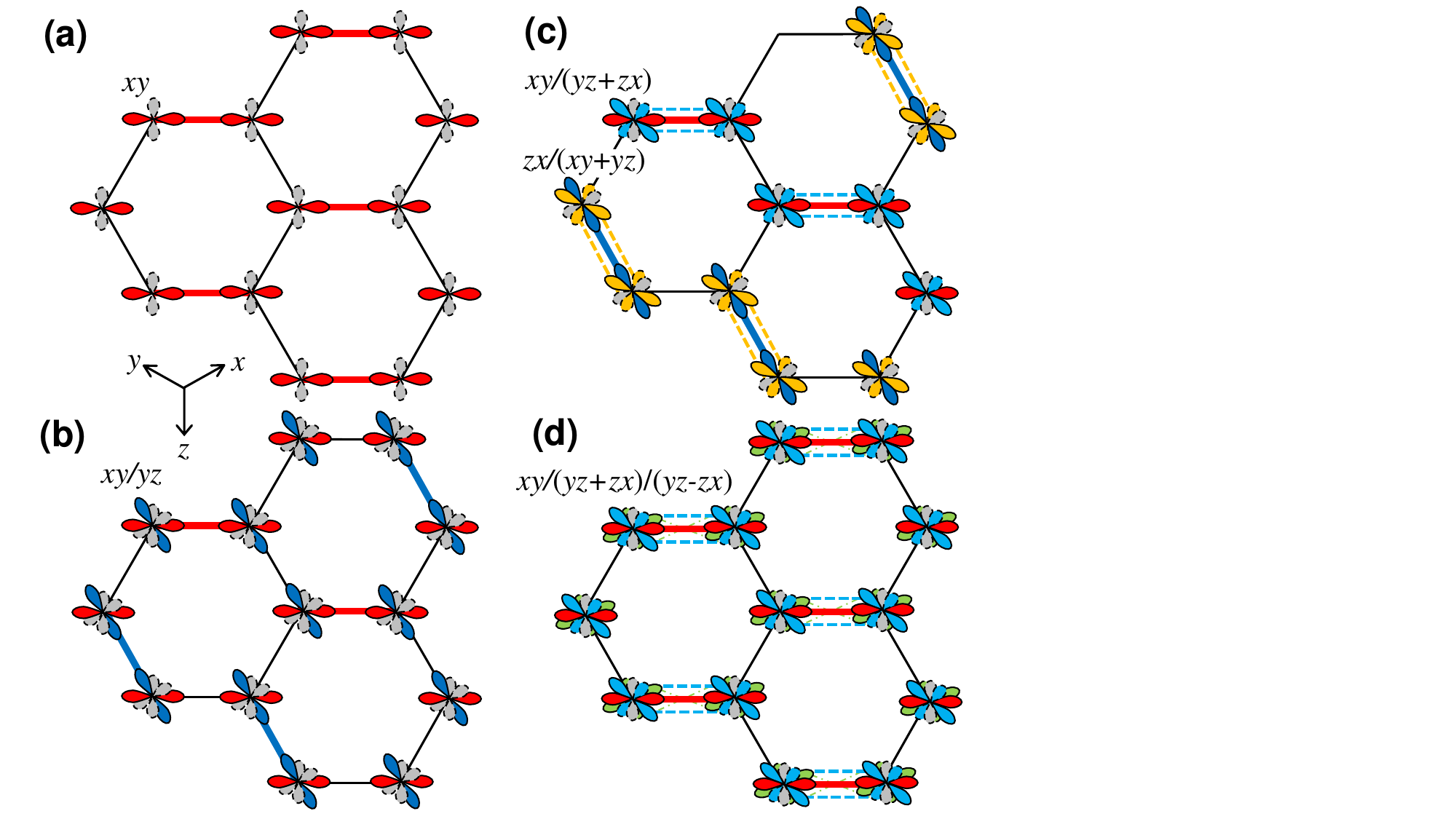}
\caption{Honeycomb network. (a) Possible orbital ordering on the $d^1$ or $d^5$ honeycomb lattice of $\alpha$-RuCl$_3$ and Na$_2$IrO$_3$ under pressure. (b) Orbital ordering with single bonding of two different active orbitals.
(c) Orbital ordering with double bonding including two orbitals at each site, case of $d^4$ in Li$_2$RuO$_3$.
(d) Orbital ordering with triple bonding on the $d^3$ honeycomb network in MoCl$_3$. The thick solid, thick dashed, and thin dot-dashed lines indicate the $\sigma$-, $\pi$, and $\delta$-bondings, respectively.}
\label{hc}
\end{figure}

Yet, it is still possible to decompose the honeycomb network into zigzag chains (the same zigzag chains are observed in quasi-1D pyroxenes). As shown in Fig.~\ref{hc}(a), assuming ferro-type orbital order of $xy$, local $xy$-$xy$ bonds can be created alternately along the zigzag chain for $d^1$ or $d^5$ systems. This can explain the fact that $\alpha$-TiCl$_3$ and $\alpha$-TiBr$_3$ (Ti$^{3+}$, $3d^1$) having the honeycomb lattice become nonmagnetic below structural transition at 217 K \cite{Ogawa1960} and 178 K \cite{Pei2020}, respectively. The direct {\it ab initio} band structure calculations show that both materials tend to dimerize \cite{Gapontsev2021} in contrast to the earlier prediction \cite{Motizuki1978}. Raman measurements for $\alpha$-TiBr$_3$ are consistent with this structural model~\cite{Pei2020}. A very similar situation is observed in $\alpha$-ZrCl$_3$ (Zr$^{3+}$, $4d^1$), where the energy gain due to dimerization prevents~\cite{Ushakov2020} realization of the $SU(4)$ spin-orbit liquid proposed theoretically~\cite{Yamada2018}. Ilmenite-type MgVO$_3$ (V$^{4+}$, $3d^1$) also exhibits V-V dimers on the honeycomb lattice below 500 K \cite{Yamamoto2022}. The transition temperature increases up to 625 K in ZnVO$_3$ where the Zn-Zn dimers coexist with the V-V dimers \cite{Yamamoto2024}.

4$d$ and 5$d$ honeycomb systems with $d^1$ or $d^5$ configurations (such as Na$_2$IrO$_3$, $\alpha$-RuCl$_3$, and $\alpha$-RuBr$_3$) do not show dimerization at ambient pressure, which is probably suppressed by the strong spin-orbit coupling \cite{Foyevtsova2013}. Under high pressure, $\alpha$-RuCl$_3$ and $\alpha$-RuBr$_3$ exhibit Ru-Ru dimerization with the ferro-type orbital order as illustrated in Fig.~\ref{hc}(a) \cite{Bastien2018,Shen2024}. Iridates such as $\alpha$-Li$_2$IrO$_3$\cite{Hermann2018}, Ag$_3$LiIr$_2$O$_6$\cite{Jin2024a}, and Na$_2$IrO$_3$\cite{Hu2018,Xu2023} also undergo a pressure induced phase transition to the dimerized state. Since the direct hopping terms $t_{xy}$, $t_{yz}$, and $t_{zx}$ are relatively large in the 4$d$ and 5$d$ systems, the energy gain by the dimerization overcomes the spin-orbit coupling when they are enhanced under pressure. One of the $t_{2g}$ orbitals is selected to form the anisotropic dimers as illustrated in Fig.~\ref{hc}(a).

In this context, it is highly interesting that Li$_2$RuO$_3$ with Ru$^{4+}$ ($4d^4$, two holes in $t_{2g}$) shows a very different packing of dimers: instead of the parallel dimers realized in all previously discussed materials, the armchair pattern shown in Fig.~\ref{hc}(b) appears below $T_s = 540$~K, where structural and metal-insulator transitions occur simultaneously~\cite{Miura2007,Miura2009}.

Assuming ferro-type orbital order of $xy$ and $yz$ holes in this $d^4$ electronic configuration, $xy$-$xy$ and $yz$-$yz$ singlet bonds (antibonding states in the case of holes) can be created as illustrated in Fig.~\ref{hc}(b). The difference of dimer arrangement between Figs.~\ref{hc}(a) and (b) is due to the ``$\pi$ phase shift'' of the dimerization between the neighboring zig-zag chains. With the Li ion at the center of the hexagon, the dimer arrangement of Fig.~\ref{hc}(b) would be favored since otherwise the hexagon needs to be compressed along the horizontal direction in the case of Fig.~\ref{hc}(a).

Interestingly, in LiVO$_2$, two electrons at each site participate in two singlet bonds on different metal-metal bonds, leading to the formation of V$_3$ trimers with $S_{\text{tot}}=0$. In Li$_2$RuO$_3$, the situation seems similar, but with holes instead of electrons. However, these holes are involved in the formation of only one short bond with the same neighbor; no larger cluster, such as short hexagons, appears. Instead, the $\frac{1}{\sqrt{2}}(yz+zx)$ orbitals form $\pi$-bonding mainly through $d$-$p$-$d$ hoppings illustrated in Fig.~\ref{1D-orbitals} in addition to the $xy$-$xy$ $\sigma$-bonding \cite{Kimber2014}. Under the orbital order with $yz$, $\frac{1}{\sqrt{2}}(zx+xy)$ on one Ru-Ru bond and $xy$, $\frac{1}{\sqrt{2}}(yz+zx)$ on another, the short bonds are further stabilized by the double bonding in the nonmagnetic phase of Li$_2$RuO$_3$ [Fig.~\ref{hc}(c)]. 

This makes Ru ions in dimers even more strongly coupled. Indeed, the short (long) Ru-Ru bond length is about 2.6\AA~(3.0\AA) in Li$_2$RuO$_3$ while the short (long) Ir-Ir bond length is about 2.7\AA~(3.0\AA) in the pressure induced dimerized phase of $\alpha$-Li$_2$IrO$_3$ with Ir$^{4+}$ ($5d^5$, only $\sigma$-bonding)~\cite{Hermann2018}. The short Ru-Ru bond would be consistent with the armchair arrangement of the dimers. A recent optical study reports that the relatively weak $\pi$-bonding by the $yz$ and $zx$ orbitals can be selectively broken and the partially disordered phase can be optically created \cite{McArdle2022}. In a similar manner, TcCl$_3$ with Tc$^{3+}$ ($4d^4$, two holes in $t_{2g}$) has Tc-Tc dimers on the honeycomb lattice, but the dimer pattern is different from that of Li$_2$RuO$_3$: they are parallel to each other \cite{Poineau2012} (presumably due to the absence of ions sitting in the centers of hexagons as theoretically analyzed by Jackeli and Khomskii \cite{Jackeli2008}).

$\alpha$-MoCl$_3$ with Mo$^{3+}$ ($4d^3$, three electrons in $t_{2g}$) undergoes a magnetic-nonmagnetic transition around $T_s=585$ K with the strong Mo-Mo dimerization below $T_s$\cite{Schafer1967,Hillebrecht1997,McGuire2017}. The Mo-Mo dimerization in the honeycomb network is illustrated in Fig.~\ref{hc}(d).  In addition to the $xy$-$xy$ $\sigma$-bonding and the $\frac{1}{\sqrt{2}}(yz+zx)$ $\pi$-bonding, $\frac{1}{\sqrt{2}}(yz-zx)$ orbitals may form $\delta$-bonding although $\delta$-bonding is usually weak. Indeed, the magnetic susceptibility drops below $T_s$~\cite{McGuire2017}, so that all three $t_{2g}$ electrons seem to be bonded. Interestingly, SrRu$_2$O$_6$~\cite{Hiley2014}, BaRu$_2$O$_6$~\cite{Marchandier2020}, and AgRuO$_3$~\cite{Prasad2017}, which share the same honeycomb lattice and the Ru$^{5+}$ $4d^3$ electronic configuration, do not dimerize; instead, at least some of them appear to exhibit quasimolecular orbitals on Ru$_6$ hexagons~\cite{Streltsov2015,Schnelle2021}.

\hfill \break {\bf \noindent THREE-DIMENSIONAL LATTICES\\}
In the spinel-type $AM_2X_4$ compounds, which crystal structure is shown in Fig.~\ref{structure}(e), the $M$ sites occupied by transition metals form the pyrochlore network in which $M_4$ tetrahedra share their corners. $MX_6$ octahedra again share their edges, and, e.g., along the [1,$\pm$1,0] direction only the transfer integral between the $xy$ orbitals are allowed by symmetry, if only a direct $t_{2g}-t_{2g}$ hoppings considered (effect of the ligand assisted processes will be discussed later). Also along the [1,0,$\pm$1] and [0,1,$\pm$1] directions, only the $zx$ and $yz$ orbitals have non-zero transfer integrals, respectively. Consequently, the $xy$, $yz$, and $zx$ orbitals form 1D bands along the [1,$\pm$1,0], [1,0,$\pm$1], or [0,1,$\pm$1] directions, respectively. Similar to the argument for the triangular lattice, symmetry breaking of the $t_{2g}$ orbital degeneracy due to band or local Jahn-Teller effect can create quasi-1D Fermi surfaces driving the Peierls instability. In this section, transition metal compounds with 3D pyrochlore network are discussed in the context of orbital-induced Peierls picture.

\hfill \break {\bf \noindent CuIr$_2$S$_4$ and LiRh$_2$O$_4$.}
There are 5.5 $t_{2g}$ electrons (0.5 holes) per transition metal in these materials. In the cubic spinel, the 0.5 holes are distributed in the three $t_{2g}$ bands and each band accommodates 1/6 holes. The six pairs of 1D Fermi surfaces (six pairs of parallel planes in the 3D $k$-space) may have the Peierls instability due to electron-phonon interaction or electron-electron interaction. Since the paired parallel planes are spanned by the wave vectors along [1,$\pm$1,0], [1,0,$\pm$1], [0,1,$\pm$1] with the magnitude of $\pi/6$ (unit of the wave vectors is $1/a$, where $a$ is the $M$-$M$ distance), the pyrochlore network may undergo distortion with 12 times periodicity along [1,$\pm$1,0], [1,0,$\pm$1], and [0,1,$\pm$1] directions. However, as discussed for the triangular lattice case, the indirect $M$-$M$ transfers via the ligands are not negligibly small. The finite transfer integrals between the different $t_{2g}$ orbitals deform the Fermi surfaces and break the nesting condition. Indeed, the calculated Fermi surfaces for the cubic CuIr$_2$S$_4$ have no Fermi surface nesting~\cite{Oda1995,Sasaki2004,Sarkar2009}. 

\begin{figure}[t!]
\centering
\includegraphics[width=0.55\textwidth]{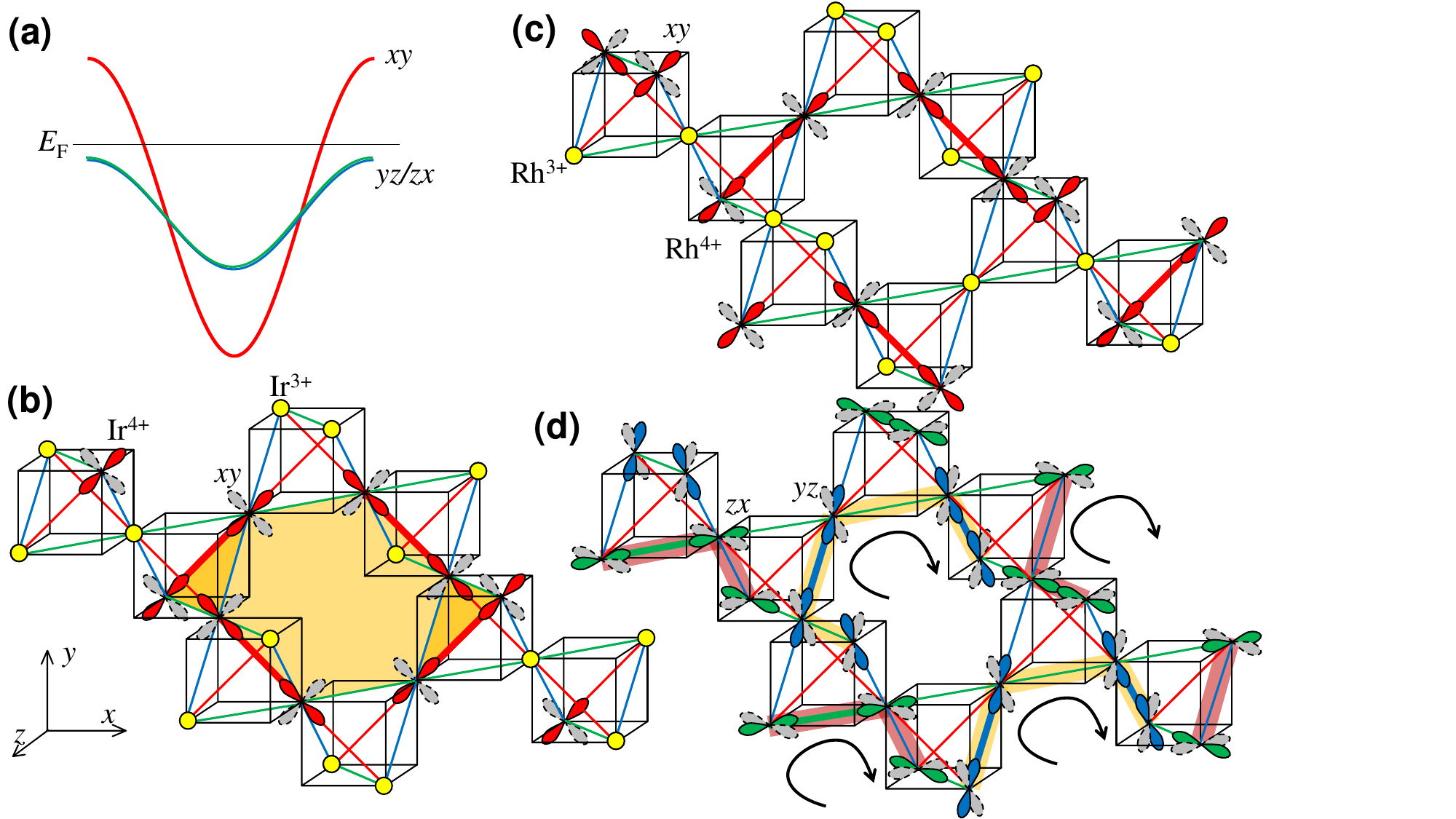}
\caption{(a) Sketch of $yz$, $zx$, and $xy$ bands under band Jahn-Teller distortion (elongated along the $c$- or $z$-axis) for CuIr$_2$S$_4$ and LiRh$_2$O$_4$. (b) Charge and orbital ordering with the octamer (shaded) on the Ir pyrochlore network in CuIr$_2$S$_4$. Ir$^{3+}$ with completely filled $t_{2g}$ shell are shown by yellow balls, the Ir 5$d$ $xy$ orbitals of Ir$^{4+}$ form the short Ir-Ir bonds along [1,$\pm$1,0], which are indicated by the thick lines. (c) Charge and orbital ordering on the Rh pyrochlore network in LiRh$_2$O$_4$. The Rh 4$d$ $xy$ orbitals of the Rh$^{4+}$ site form the short Rh-Rh bonds along [1,$\pm$1,0]. (d) Orbital ordering and dimerization in MgTi$_2$O$_4$. The Ti 3$d$ $yz$ and $zx$ orbitals form the short Ti-Ti bonds along [0,1,$\pm$1] and [1,0,$\pm$1], respectively. The helices formed by the short and long Ti-Ti bonds are shaded and their helicity are indicated by the arrows. Only transition metals are shown by yellow balls or sites with orbitals}
\label{pyro}
\end{figure}

In the orbital-induced Peierls scenario, one of the $t_{2g}$ bands, e.g. $xy$, becomes wider than the other two (due to band Jahn-Teller effect) and accommodates all the holes as illustrated in Fig.~\ref{pyro}(a). Then the quasi-one-dimensional band with 3/4-filling can provide the Peierls instability with four times periodicity along [1,$\pm$1,0].

CuIr$_2$S$_4$ exhibits a metal-insulator transition around 230 K \cite{Nagata1994} with tetragonal elongation ($a<c$) and trigonal octamer charge ordering \cite{Radaelli2002,Ohashi2025} as shown in Fig.~\ref{pyro}(b). All the holes are accommodated in the $xy$ orbitals and the
Ir$^{3+}$-Ir$^{3+}$-Ir$^{4+}$-Ir$^{4+}$ arrangement with the Ir$^{4+}$-Ir$^{4+}$ dimers is formed along [1,$\pm$1,0] in agreement with the orbital-induced Peierls scenario. CuIr$_2$S$_4$ with the unique electronic and lattice properties harbor the partially disordered state induced by light or X-ray illumination \cite{Ishibashi2002,Takubo2005,Kiryukhin2006,Naseska2021}, where the long range charge order is destructed but the tetragonal distortion (and orbital order) remains. 

LiRh$_2$O$_4$ undergoes the cubic to tetragonal transition (again $a<c$) at 220 K which is followed by the tetragonal charge ordering and the Rh$^{4+}$-Rh$^{4+}$ dimerization as shown in Fig.~\ref{pyro}(c). The Rh$^{3+}$-Rh$^{4+}$ charge fluctuation \cite{Nakatsu2011} and the local Rh$^{4+}$-Rh$^{4+}$ dimers \cite{Shiomi2022} are observed between 220 K and 170 K, indicating that orbital symmetry breaking plays more important roles in LiRh$_2$O$_4$. Here, one can speculate that, in the Rh 4$d$ bands narrower than the Ir 5$d$ bands, the band Jahn-Teller distortion alone can stabilize this particular distortion. In case of CuIr$_2$S$_4$, the local Ir-Ir dimers are found even in the cubic phase \cite{Bozin2019} while the Rh-Rh dimers are observed only in the tetragonal phase. The charge ordering of LiRh$_2$O$_4$ satisfies the Anderson condition on the pyrochlore network indicating importance of the Coulomb repulsive force between holes while the octamer charge ordering of CuIr$_2$S$_4$ violates it. In LiRh$_2$O$_4$, the energy gain by the dimerization is smaller due to the electron correlation while the band Jahn-Teller instability is stronger than in CuIr$_2$S$_4$.

\hfill \break {\bf \noindent MgTi$_2$O$_4$.}
Spinel-type MgTi$_2$O$_4$ (Ti$^{3+}$, $3d^1$) exhibits a metal-insulator transition around 260 K which is accompanied by a structural transition from cubic to tetragonal ($a>c$) \cite{Isobe2002a}. The tetragonal structure has Ti$^{3+}$-Ti$^{3+}$ dimerization with spirals of long and short Ti-Ti bonds \cite{Schmidt2004}. With $a>c$, the $yz$ and $zx$ bands are wider than the $xy$ band and accommodate the Ti 3$d$ electrons. Quasi one-dimensional $yz$ and $zx$ bands are formed along [0,1,$\pm$1] and [1,0,$\pm$1] and accommodate 0.5 electron respectively. The 1/4-filled $yz$ and $zx$ bands provide the Peierls instability with four time periodicity of $yz$-$yz$ and $zx$-$zx$ dimers as shown in Fig.~\ref{pyro}(d). 
However, since the TiO$_6$ octahedron is also compressed along the $c$-axis, the center of mass of the $xy$ band becomes lower in energy than those of the $yz$ and $zx$ bands due to the ligand field. Thus MgTi$_2$O$_4$ shares the competition between the band Jahn-Teller effect and the ligand field effect (i.e. local Jahn-Teller effect for which change of bandwidth is not important) with NaTiO$_2$. Similar to Na$_2$Ti$_3$Cl$_8$, it is possible to explain the orbital ordering of MgTi$_2$O$_4$ based on the localized picture \cite{DiMatteo2004} although it is difficult to describe the metal-insulator transition. The localized nature of the Ti-Ti dimer has experimentally been suggested from the survival of the local Ti-Ti dimers above the transition temperature \cite{Yang2020} and the multiplet structure of the Ti 2$p$ X-ray absorption spectrum \cite{Yamaguchi2022}. Both the itinerant model and the localized model provide the same conclusion on the orbital ordering of MgTi$_2$O$_4$.

\hfill \break {\bf \noindent AlV$_2$O$_4$, GaV$_2$O$_4$, and LiV$_2$O$_4$.}
Different structural models have been proposed for the spinel-type AlV$_2$O$_4$ (V$^{2.5+}$, $3d^{2.5}$)~\cite{Matsuno2003,Horibe2006,Browne2017}. A recent structural study shows that AlV$_2$O$_4$ appears to exhibit a remarkable formation of two distinct types of clusters: V$^{2+}_4$ tetrahedra and V$^{3+}_3$ triangles along with a single unpaired V$^{3+}$ ion with a localized $S=1$, i.e. it can be considered as Al$_4$($[\text V_4]^{8+}$$[\text V_3]^{9+}$$\text V^{3+}$)O$_{16}$ \cite{Browne2017}.
The V$^{2+}$/V$^{3+}$ charge disproportionation is confirmed by X-ray photoemission spectroscopy \cite{Okawa2024}.

\begin{figure}[t!]
\centering
\includegraphics[width=0.48\textwidth]{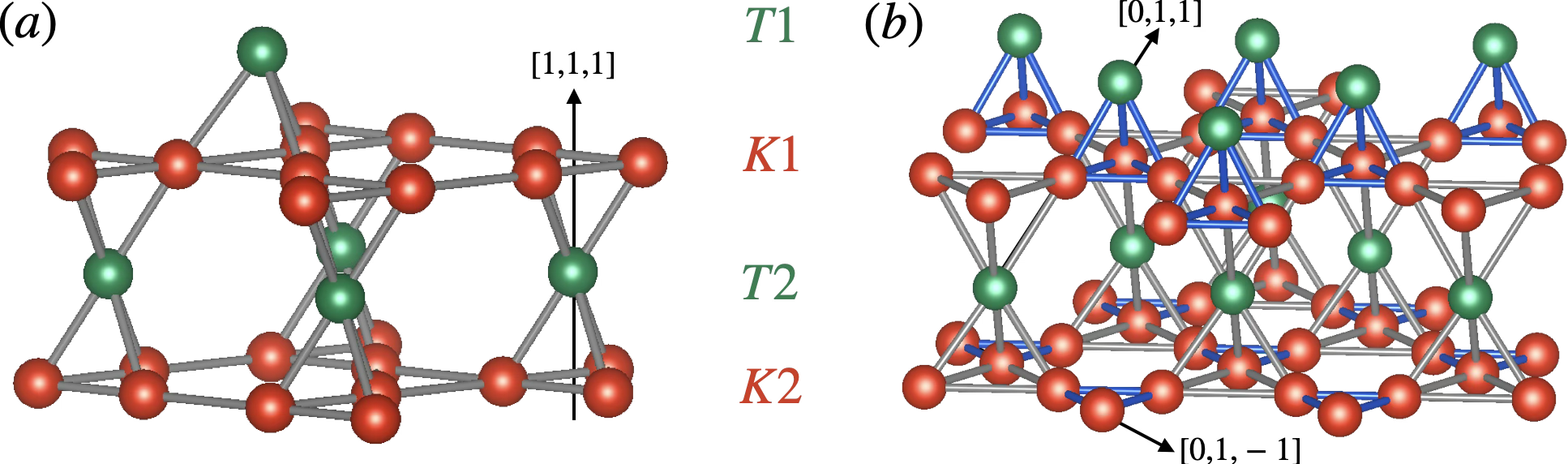}
\caption{(a) High-temperature structure of spinel AlV$_2$O$_4$, which can be decomposed into kagome (red spheres) and triangular (green spheres) layers of V ions. (b) A possible low-temperature structure. Short V--V bonds, shown in blue, illustrate the formation of V$_4$ tetrahedra and V$_3$ triangles. }
\label{AlV2O4-structure}
\end{figure}

Spinel lattice can be represented as alternating along the [1,1,1] direction between kagome and triangular layers of transition-metal ions, see Fig. \ref{AlV2O4-structure}(a). The V$^{3+}$ ions with a $d^2$ configuration in one of the kagome planes (K1) group into trimers. In the next kagome layer (K2), a V$^{2+}$ ion with one additional electron forms tetrahedra with V$^{3+}$ ions in the neighboring triangular plane (T1). The remaining V$^{3+}$ ion, with a dangling $S=1$, resides in the last triangular layer (T2).

\begin{figure}[t!]
\centering
\includegraphics[width=0.65\textwidth]{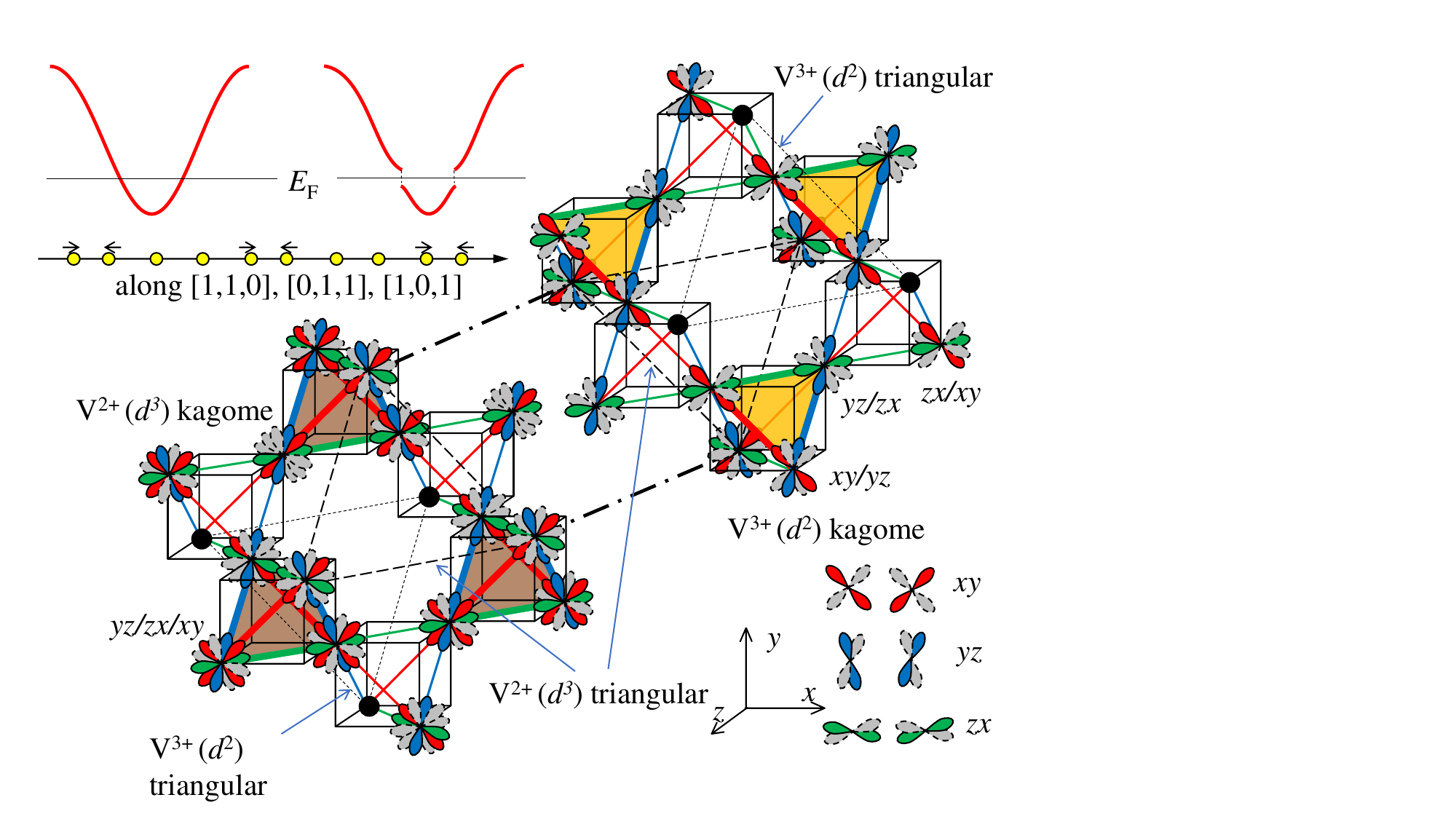}
\caption{1/4-filled bands along the [1,1,0], [0,1,1], and [1,0,1] chains and V$^{2+}$/V$^{3+}$ charge ordering for AlV$_2$O$_4$. The black circles indicate \textcolor{cyan}{V$^{3+}$ sites in triangular layers}. In the V$^{3+}$ ($d^2$) kagome network layer, the V$^{3+}$ trimers are formed by the $yz$/$zx$, $zx$/$xy$, and $yz$/$xy$ orbital ordering. The shaded triangles and tetrahedra indicate the V trimers and tetramers, respectively. In the V$^{2+}$ ($d^3$) kagome network layer, the tetramers are formed with the V$^{2+}$ sites ($yz$/$zx$/$xy$) in the neighboring triangular lattice layer. The dash-dotted lines indicate the same V$^{2+}$ site.}
\label{AlV2O4}
\end{figure}

Interestingly, if one assumes that all V$_4$ tetrahedra are oriented in the same direction as in Fig.~\ref{AlV2O4-structure}(b), this unusual coexistence of tetrahedra and trimers can be explained by the orbital-induced Peierls transition. 
Along the [1,-1,0], [1,0,-1], and [0,1,-1] chains in the kagome layers, a doubling of the period occurs due to the two half-filled $t_{2g}$ bands at each V site, similarly to Na$_2$Ti$_3$Cl$_8$ (see Fig.~\ref{AlV2O4}). 
However, the average valence of V is 2.5+, leaving additional 0.5 electron per site. 
This corresponds to a 1/4 filling of the remaining $t_{2g}$ orbitals, which in turn leads to a tetramerization (period 4) along the [1,1,0], [1,0,1], and [0,1,1] chains that cross both the kagome and triangular layers.

This picture of the orbital-induced Peierls transition explains the formation of isolated V ions, trimers, and tetrahedra all pointing in the same direction (i.e., a ferroelectric structure, since it breaks inversion symmetry). 
However, available experimental data indicate a disorder in the orientation of the tetrahedra~\cite{Browne2017}, which may arise from small off-stoichiometry and/or partial cation inversion between Al and V sites - a common feature in spinels, see e.g. \cite{goodenough1963,zhidkov2021}. 
Another important factor is configurational entropy, given that the transition to the clusterized structure occurs at a very high temperature (700~K). 
Although this prevents a definitive assignment of the transition mechanism, orbital degrees of freedom are clearly central - whether through local chemical bonding between V ions or via cooperative effects like the Peierls transition. Notably, as we argue in the ``Peierls or not really Peierls'' section above, these two possibilities are not mutually exclusive and often go hand in hand.

Similar to AlV$_2$O$_4$, spinel-type GaV$_2$O$_4$ (V$^{2.5+}$, $3d^{2.5}$) exhibits the trimer-tetramer structure \cite{Browne2018} in which mixed valence state can be realized by Zn substitution for Ga \cite{Browne2020}. Ga$_{1-x}$Zn$_x$V$_2$O$_4$ with $0.06 \le x \le 0.75$ still have disordered trimers and/or tetramers although the global structure is cubic. Most probably, while V$^{2+}$ tetramer can be decomposed into trimer and monomer with the valence change from V$^{2+}$ to V$^{3+}$, the V$^{3+}$ trimers in the kagome network layers are robust.

The orbital-induced Pierels mechanism would be relevant for the pressure induced insulating phase of LiV$_2$O$_4$~\cite{Kondo1997,Urano2000,Anisimov1999,Eyert1999,Uehara2015}. The average valence of V is +3.5 ($3d^{1.5}$) in LiV$_2$O$_4$. Assuming that the V sites in the triangular lattice layers are V$^{5+}$ ($d^0$), the average valence of V of the kagome network layers are V$^{3+}$. Robustness of the trimers in the V$^{3+}$ kagome layers would be consistent with the pressure induced insulating phase of LiV$_2$O$_4$.

\hfill \break {\bf \noindent CALCULATION AND EXPERIMENTAL   METHODS\\}
\hfill  {\bf \noindent Theoretical methods.} Equation \eqref{chi0} has been widely used to study the Peierls transition. It not only allows one to find a divergence at a certain $q$-vector, suggesting a new distorted structure, but also to determine which bands (or orbitals) are involved and thereby to gain insight into the microscopic mechanism behind the transition. However, \textcolor{brown}{first of all,} it should be noted that \eqref{chi0} represents an expression for the generalized susceptibility of the electronic system, and an instability can occur not necessarily in the charge (lattice) sector; a divergence in $\chi_0(\vec q)$ can also signal a tendency to form, e.g., a long-range magnetic order or to lead to superconductivity.

Second, large peak in the Lindhard susceptibility \eqref{chi0} alone does not guarantee a Peierls instability. The full criterion involves the dimensionless electron-phonon coupling $\lambda_{\vec q}$. Following the standard Dyson equation formalism for phonons, see, e.g., \cite{Pouget2016}, Eq. (20), it can be written as
\begin{eqnarray}
\lambda_{\vec q}= \frac{2|g_{\vec q}|^2}{M\omega_{0}(\vec q)} \chi_0(\vec q),     
\end{eqnarray}
where $\omega_0(\vec q)$ is the bare phonon frequency, $M$ is the ionic mass (or the relevant reduced mass for the phonon mode), and $g_{\vec q}$ is the electron-phonon coupling matrix element. An instability requires $\lambda_{\vec q} > 1$. Hence, a sharp (logarithmic) divergence in $\chi_0(\vec q)$ can be suppressed by weak coupling or a stiff lattice, while a soft phonon can stabilize the transition even with ``moderate'' nesting. This underscores the importance of lattice softness, which is naturally captured by {\it ab initio} phonon calculations. Therefore, a more direct approach is to perform DFT calculations of the phonon spectrum of the high-temperature structure. The phonon softening, or the appearance of imaginary phonon modes at a specific wave vector $\vec q$, is strong evidence of a lattice instability. While such standard calculations do not explicitly compute the electron-phonon matrix elements, they directly capture the bare phonon frequency $\omega_0(\vec q)$ - the lattice softness -which critically enhances the dimensionless coupling constant $\lambda_{\vec q}$ that drives the Peierls transition.

Another important question is which approximation should be used to study possible nesting and instability in \eqref{chi0}. Typically, one considers nonmagnetic DFT calculations, but interactions can do more than just renormalize the electronic spectrum close to the Fermi level. For example, spin-orbit coupling can substantially modify the electronic structure of materials based on heavy elements and therefore must be taken into account (see, e.g., the case of CsW$_2$O$_6$~\cite{Streltsov2016}). And even more delicate question is whether strong Coulomb correlations should be taken into consideration in a static way, as is done in the DFT+U approximation, or whether methods such as dynamical mean-field theory (DMFT)~\cite{Georges1996} with a frequency-dependent self-energy (which properly renormalizes the electronic structure close to $E_F$) must be used. Interestingly, pulling localized states away from the Fermi level using a static potential, as in DFT+U, sometimes helps to reveal nesting in materials subjected to Peierls transitions, as occurs, e.g., in the hollandite K$_2$Cr$_8$O$_{16}$~\cite{Toriyama2011}.

Phonon calculations and crystal structure optimization are also not always as straightforward as anticipated. On the one hand, {\it ab initio} methods are generally thought to provide correct crystal structures only if the ground-state electronic structure is described properly. Therefore, one might expect that including the Hubbard terms is essential for transition metal oxides. However, DFT+U often fails to describe the Peierls transition; for example, in canonical VO$_2$, it yields uniform V chains~\cite{Wickramaratne2019}. This occurs because DFT+U replaces the large energy gain from the formation of molecular orbitals - which is proportional to the hopping $t$ - with a much smaller exchange coupling of the order $t^2/U$. Extensions of DMFT taking into account both (local) correlation effects and band formation on the same footing are promising in describing materials with Peierls transition~\cite{Mlkvik2024}.

\hfill \break {\bf \noindent Experimental methods.} It is worth noting that historically, the formation of clusters was often detected by thermodynamic measurements such as specific heat signaling phase transitions, or by magnetic susceptibility, since clusters typically turn out to be nonmagnetic. The most direct and significant technique for studying Peierls distortions is, of course, diffraction. However, this approach has limitations not only related to issues such as twinning, incoherent scattering, etc., but also to problems typical for materials undergoing Peierls transition. As discussed above, these transitions often occur in two-dimensional systems with triangular or kagome lattices. Such layered materials frequently exhibit stacking faults, i.e., irregularities or disruptions in the ideal sequence of atomic or layer stacking. This arises from the weak (often van der Waals) bonding between layers. For example, this seems to be precisely the problem in LiVO$_2$~\cite{Kojima2019,Kojima2023}, where magnetic susceptibility shows a drop~\cite{Kobayashi1969} most probably associated with the formation of nonmagnetic V trimers, yet diffraction techniques (and even Raman measurements~\cite{Ponosov2024}) are unable to refine structural trimers in the low-temperature phase. 

The orbital anisotropy and the electronic configuration of particular ions can be probed by X-ray absorption spectroscopy, especially when combined with theoretical many-electron  calculations, which are typically performed nowadays within the configuration interaction approximation. This combination can reproduce very fine details of X-ray absorption spectra and extract the electronic wave functions that fully characterize the system. Such an approach was used, for example, to demonstrate that the dimerization in VO$_2$ cannot be described within a conventional Peierls theory; electronic correlations are involved in reducing charge fluctuations and making the system more 1D (via orbital degrees of freedom) and thus susceptible to a Peierls-like transition~\cite{Haverkort2005}.

The charge modulation/disproportionation and valence state can be probed by core level photoemission spectroscopy. In particular, hard X-ray photoemission spectroscopy (HAXPES) can detect charge fluctuations and the degree of electron localization within clusters formed in the low-temperature phase~\cite{Streltsov2025}.

Another extremely important experimental technique, which has been discussed in the Peierls or not really Peierls section is PDF analysis of neutron or X-ray diffraction data. It allows to figure out what happens with clusters in the high-temperature phase, i.e. above $T_s$. One of the main results obtained in different PDF studies is that clusters typically survive at temperatures much higher than $T_s$, see Table~\ref{Tab:PDF}.

\begin{table}[t!]
\begin{tabular}{lcc}
\hline
\hline 
Material & Structural transition & Clusters exist at least up to\\
\hline
LiRh$_2$O$_4$ & 170 K & 350 K~\cite{Knox2013} \\
NaTiSi$_2$O$_6$ & 210 K & 490 K \cite{Koch2021} \\
CuIr$_2$S$_4$ & 230 K & 780 K$^*$ \cite{Bozin2019} \\
MgTi$_2$O$_4$ &  260 K & 500 K \cite{Yang2020} \\
GaV$_2$O$_4$ &  415 K & 1100 K \cite{Browne2018} \\
Li$_2$RuO$_3$ &  540 K & 920 K \cite{Kimber2014} \\
AlV$_2$O$_4$ &  700 K & 1100 K \cite{Browne2017} \\
\hline
\hline
\end{tabular}
\caption{List of materials for which PDF study has been performed. In last column the highest experimentally studied temperature at which PDF observes short metel-metal bond is indicated (this means that they can survive at even higher $T$). $^*$ is used to indicate vanishing difference between short and long bonds at high $T$.
}
\label{Tab:PDF}
\end{table}

For pyroxene NaTiSi$_2$O$_6$, where dimerization of quasi-1D chains occurs at $T_s = 210$K, PDF finds that $\sim$ 6 Ti sites remain correlated at 490K.\cite{Koch2021} I.e. we see that while a long-range order of dimers disappear there still exist pieces of dimerized chains (3 dimers in average). Difference between short ($l_s$) and long ($l_l$) Ti-Ti bond length is $\delta l = l_l - l_s= 0.154$\AA~in low-temperature phase \cite{Redhammer2003}, reduces to 0.124\AA~at 490K \cite{Koch2021}.

\begin{figure}[t!]
\centering
\includegraphics[width=0.48\textwidth]{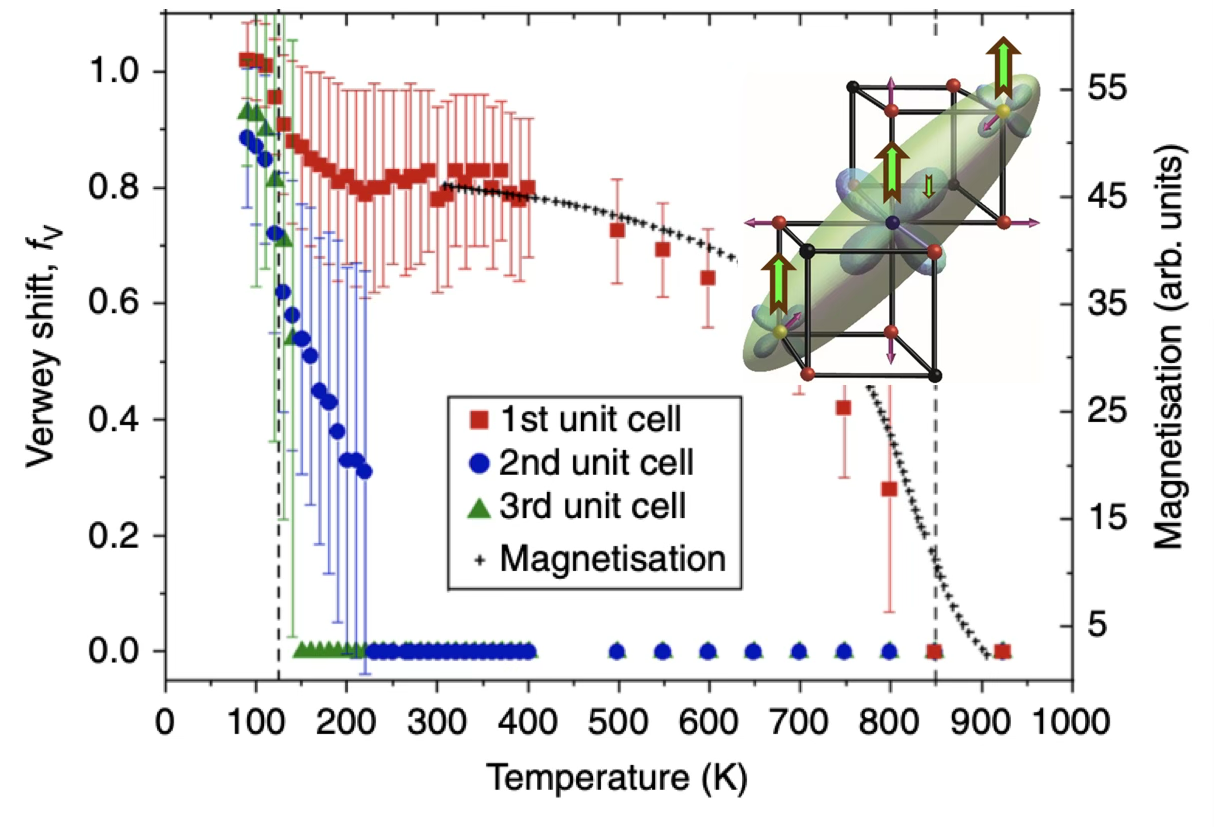}
\caption{Shift parameter $f_v$, which characterizes the strength of distortions leading to the formation of trimerons, plotted together with the temperature dependence of the magnetization in Fe$_3$O$_4$. The long-range order of trimerons breaks at the Verwey temperature $T_V \approx 125$~K (structural transition; green symbols), whereas local order (blue and red symbols) persists up to much higher temperatures. The inset shows a Fe$^{3+}$-Fe$^{2+}$-Fe$^{3+}$ trimeron. Adapted from~\cite{perversi2019}.} 
\label{Fe3O4-structure}
\end{figure}

Another example with clusters is magnetite, Fe$_3$O$_4$, which adopts the spinel structure with edge-sharing FeO$_6$ octahedra. Below the Verwey transition at $T_V \approx 125$~K, three Fe ions form linear clusters of Fe$^{3+}$--Fe$^{2+}$--Fe$^{3+}$ trimerons~\cite{Senn2012}. X-ray PDF studies demonstrate that the long-range trimeron order disappears just above $T_V$, while short-range correlations persist up to the Curie temperature $T_C = 850$~K and its temperature dependence exactly copies that of the magnetization, see Fig.~\ref{Fe3O4-structure} and \cite{perversi2019}.

This 20\% decrease of $\delta l$ in the two phases of NaTiSi$_2$O$_6$ contrasts a lot with what is observed in CuIr$_2$S$_4$~\cite{Bozin2019}. In this material, short Ir-Ir dimers have $l_s =2.97$\AA~($\delta l = 0.697$\AA) in the low-temperature phase~\cite{Ohashi2025}. There are still ``short'' in some sense Ir-Ir bonds above the structural transition, but they become 3.5\AA~and the corresponding short-long bond difference $\delta l$ decreases down to  0.08\AA~\cite{Bozin2019}, i.e. on nearly 90\%. Such a strong modification of the local structure implies a change of the orbital state between the two phases of CuIr$_2$S$_4$~\cite{Bozin2019}.

Thus, there can be at least three situations. First, the clusters stay nearly unchanged and start to ``float'' over the lattice in the high-temperature phase. Moreover, some pieces of the initial dimerized structure and the corresponding orbital state can remain locally intact. It would be very interesting to study in more detail how the disorder develops with temperature in this situation. The second option is when the bond lengths in clusters change considerably, as, e.g., in CuIr$_2$S$_4$. Most probably, in this case we have rich multiplet structure in the cluster and an interplay between different degrees of freedom. Thus, volume expansion (or thermal activation) changes the electronic state considerably by (i) redistributing electrons among the molecular and localized orbitals in the cluster, (ii) transforming their wavefunctions, e.g. going from the MO-LCAO-like to the HL-like state, or (iii) even activating other multiplets. The last option is complete destruction of clusters immediately above $T_s$. This situation appears rather rare. A single documented case from PDF analysis is IrTe$_2$~\cite{Yu2018a,Yu2018b}, while extended X-ray absorption fine structure (EXAFS) seems to suggest locally persisting Ir-Ir dimers above $T_s$~\cite{Paris2016}.

In electrochemistry the scanning transmission electron microscopy is routinely used for direct visualization of atomic structure. It was proven to be extremely helpful in studying formation of clusters, disorder effects, see, e.g., \cite{savina2022}, and can supplement PDF analysis.

Finally, nuclear magnetic resonance (NMR) is sensitive to the local symmetry and the occupation of valence $d$-orbitals since they strongly affect the electric field gradient (EFG) tensor at the nucleus. The NMR shift typically clearly signals the transition to the nonmagnetic state that often accompanies Peierls transitions~\cite{Shimizu2008,Arapova2017,Heinmaa2022}. Moreover, besides detecting Peierls transitions, NMR  can also provide information about the character of involved orbitals via studying anisotropic tensors of Knight shift and quadrupole splitting frequency, see e.g. \cite{Jinno2013}.

\hfill \break {\bf \noindent RELATED EFFECTS\\} 
{\bf \noindent Pressure effect.} This is a rather typical situation for Mott insulators: external pressure increases the overlap between $d$ orbitals and therefore their bandwidth $W$, while the on-site Coulomb repulsion (Hubbard $U$) decreases due to increased screening. Consequently, the $W/U$ ratio grows, and pressure typically lowers the temperature of the metal-insulator transition~\cite{Khomskii2014}.

In clusterized materials, the situation can be just the opposite. The formation of clusters often leads to reduced volume, and such states can therefore be stabilized by pressure. E.g. chemical pressure due to  substitution of Na by a smaller Li ion increases temperature of structural transition in LiTiSi$_2$O$_6$ with respect to NaTiSi$_2$O$_6$~\cite{Isobe2002}. The whole system remains insulating in this case, while electrons can fluctuate between sites within the cluster, like it happens e.g. in LiZn$_2$Mo$_3$O$_8$~\cite{Streltsov2025,Nikolaev2021}. Thus, one may expect very different behavior in these systems, with pressure working for the insulating phase rather than against it.

This is exactly what happens in LiV$_2$O$_4$, where pressure induces a metal-insulator transition and structural study clearly demonstrate formation of short V-V bonds under pressure\cite{Browne2020b}

However, there is another factor which should always be taken into account when we discuss pressure effect. This is the tendency to move toward a truly three-dimensional situation, characterized by a large bandwidth along all directions and suppressed nesting. These two mechanisms will counteract each other and affect the temperature of the metal-insulator transition.\\

{\bf \noindent Orbital selectivity.} Another very important aspect of the orbital-induced Peierls transition is the electronic structure of the low-temperature phase. Two very different scenarios are possible.

First, all electrons form molecular orbitals, as occurs in, e.g., LiVO$_2$, LiVS$_2$, or Na$_2$Ti$_3$Cl$_8$ with a trimerized triangular lattice. In these compounds, each V or Ti ion with a $d^2$ configuration has two neighbors in the trimer, and these two electrons participate in two molecular orbitals, see Fig.~\ref{Fig:orbital-selectivity}(a). So altogether we have 3 molecular orbitals and 6 electrons occupying all of them as shown in Fig.~\ref{Fig:orbital-selectivity}(b).

A very different situation is found in MoO$_2$ ($4d^2$) with a rutile structure, where dimers rather than trimers are formed. Due to the common-edge geometry, only one of the $t_{2g}$ orbitals can form a molecular orbital due to direct $t_{2g}-t_{2g}$ overlap. Consequently, orbital-selective behavior is anticipated, where some orbitals provide localized or metallic electrons while others participate in molecular bonding~\cite{Streltsov2014a,Streltsov2016}. Case of MoO$_2$ with two electrons per site is illustrated in Fig.~\ref{Fig:orbital-selectivity}(d). Orbital selectivity in MoO$_2$ results in the coexistence of metallic conductivity and Peierls distortions~\cite{Rogers1969,Eyert2000}. 
\textcolor{cyan}{This behaviour is seen in many systems, including cluster-Mott insulator LiZn$_2$Mo$_3$O$_8$~\cite{Streltsov2025}, high-$T_c$ superconductor 
La$_3$Ni$_2$O$_7$~\cite{Dagotto}, and others}.
Effect of orbital selectivity should be explicitly taken into account in such systems, since only part of the electronic subsystem will be able to respond to external perturbations such as a magnetic field or an electric current (this will be reflected e.g. in field and temperature behavior of the magnetic susceptibility). \\

\hfill \break {\bf \noindent FUTURE OUTLOOK\\}
In the present review, the current understandings and the research trends on the orbital-induced Peierls transitions and the related effects are discussed based on the rich literature. The multimer or cluster formation in various transition-metal compounds can be described by the orbital-induced Peierls mechanism in which the (band) Jahn-Teller instability and the Peierls instability occur in a synergetic manner. Yet there are many unsolved mysteries related to the  orbital-induced Peierls transition. In particular, the fluctuations above the transition temperature or beyond the critical pressure are insufficiently explored. Development of the experimental techniques on modern X-ray and neutron sources is expected to play vital roles to probe such fluctuations with higher energy, momentum, time, and space resolutions. It is highly challenging to theoretically describe the spin-charge-orbital-lattice fluctuations and their possible impacts on exotic superconducting states that often emerge in the proximity to the multimer states. The orbital-induced transitions often occurs near the room temperature and are accompanied by the huge entropy and lattice changes, which can be controlled by electric, magnetic, optical, and thermal stimuli. In the context of materials development, such phase transitions can be applied for future energy transformation and storage devices.

\hfill \break {\bf \noindent ACKNOWLEDGMENTS\\} 
We would like to devote this review to Daniel Khomskii our friend and teacher.

SVS thanks Paul Attfield, Simon Billinge, Emil Bozin, Martin Jansen, Klim Kugel, and Stephen Wilson  for fruitful discussions. TM thanks Naoyuki Katayama for informative discussions. SVS was supported by Ministry of Science and Higher Education of the Russian Federation through IMP UrD RAS. TM was supported by Grants-in-Aid for Scientific Research (No. JP26K00663) from Japan Society for the Promotion of Science.

\hfill \break {\bf \noindent  COMPETING INTERESTS\\} 
The authors declare no competing interests.

\end{document}